\documentclass[prl,amsmath,amssymb,showpacs,superscriptaddress,nofootinbib]{revtex4-1}
\usepackage{graphicx}
\usepackage[mathlines]{lineno}
\usepackage{dcolumn}
\usepackage{color}
\usepackage{bm}
\usepackage{overpic}
\usepackage{rotating}
\usepackage{times}
\usepackage{multirow}
\usepackage{float}
\usepackage{mathrsfs}
\usepackage{booktabs}
\usepackage[mathlines]{lineno}
\usepackage{ulem}

\input{definitions}
\begin{document}
\normalsize
\parskip=5pt plus 1pt minus 1pt
\hyphenpenalty=10000
\tolerance=1000


\title{\boldmath Observation of $\epem\to\eta\Upsilon(2S)$ and search for $\epem\to\eta\Upsilon(1S),~\gamma X_b$ at $\sqrt{s}$ near 10.75 GeV}

  \author{I.~Adachi\,\orcidlink{0000-0003-2287-0173}} 
  \author{L.~Aggarwal\,\orcidlink{0000-0002-0909-7537}} 
  \author{H.~Ahmed\,\orcidlink{0000-0003-3976-7498}} 
  \author{Y.~Ahn\,\orcidlink{0000-0001-6820-0576}} 
  \author{H.~Aihara\,\orcidlink{0000-0002-1907-5964}} 
  \author{N.~Akopov\,\orcidlink{0000-0002-4425-2096}} 
  \author{S.~Alghamdi\,\orcidlink{0000-0001-7609-112X}} 
  \author{M.~Alhakami\,\orcidlink{0000-0002-2234-8628}} 
  \author{A.~Aloisio\,\orcidlink{0000-0002-3883-6693}} 
  \author{N.~Althubiti\,\orcidlink{0000-0003-1513-0409}} 
% ...
\collaboration{The Belle II Collaboration}

\date{\today}

\begin{abstract}

We present an analysis of the processes $e^{+}e^{-}\to\eta\Upsilon(1S)$, $\eta\Upsilon(2S)$, and $\gamma X_b$ with $X_b\to\pi^+\pi^-\chi_{bJ},~\chi_{bJ}\to\gamma\Upsilon(1S)$ $(J=1,~2)$ reconstructed from $\gamma\gamma\pi^+\pi^-\ell^+\ell^-~(\ell=e,~\mu)$ final states in $19.6\rm$ \invfb of Belle II data collected at four energy points near the peak of the $\Upsilon(10753)$ resonance.
Here, $X_b$ is a hypothetical bottomonium-sector partner of the $X(3872)$.
In the process $e^{+}e^{-} \to \eta \Upsilon(2S)$, we observe a signal with a significance that exceeds $6.0\sigma$.  Based on an analysis of the Born cross-section, the hypothesis that these events are associated solely with the production of the $\Upsilon(5S)$ or $\Upsilon(10753)$ resonances is rejected at the level of 3.6$\sigma$. This suggests that the observed events are more likely to arise from a new state near the $B^{*}\bar{B}^{*}$ threshold or from a substantial continuum production.
No significant signal is observed for $e^{+}e^{-}\to\eta\Upsilon(1S)$ or $\gamma X_b$.
Upper limits on the Born cross sections for the processes $e^{+}e^{-}\to\eta\Upsilon(1S)$ and $e^{+}e^{-}\to\gamma X_b$ with $X_b\to\pi^+\pi^-\chi_{bJ}$ are determined.

\end{abstract}

\maketitle


In recent decades, a series of quarkonium-like states with unconventional characteristics have been observed at accelerator facilities~\cite{pdg}, providing compelling evidence for the existence of exotic hadronic configurations. 
Among these, bottomonium states offer particularly valuable probes for investigating the non-perturbative regime of quantum chromodynamics due to their distinctive mass scale and the relatively clean theoretical description.
In 2019, a new structure, known as $\Upsilon(10753)$, was identified in the cross-section measurements of $\epem\to\pipi\Upsilon(mS),~m=1,~2,~3$~\cite{exp_belle}. 
With more data collected from the Belle II experiment, this structure was confirmed in the measurement of the cross section of $\epem\to\omega\chi_{b1,2}$~\cite{Belle-II:2022xdi}, where its mass and width were determined with greater precision~\cite{belle2_pipiYnS}. 
The small partial width of $\Upsilon(10753)\to\omega\eta_b$~\cite{Belle-II:2023twj} suggests that it may not be exotic; however, the investigation of $\Upsilon(10753)\to\pipi\Upsilon(mS)$ and $\omega\chi_{bJ},~J=1,~2$ indicates that it is also challenging to classify $\Upsilon(10753)$ as a conventional bottomonium state~\cite{Belle-II:2022xdi,belle2_pipiYnS,Bai:2022cfz}.
Therefore, further investigation is warranted to fully understand the nature of the $\Upsilon(10753)$.

The measurement of the hadronic transition of the $\Upsilon(10753)$ with emission of an $\eta$ provides a complementary approach to understanding of this resonance.
A relatively large branching fraction of $\Upsilon(10753)\to\eta\Upsilon(nS) ~\mathrm{with}~n=1,~2$ is expected given the possibility of $4S-3D$ mixing~\cite{Li:2021jjt}.
The branching fraction for $\Upsilon(10753)\to\eta\Upsilon(1S)$ is predicted to be $(0.46-5.46)\times10^{-3}$ within this scheme~\cite{Li:2021jjt}, which is comparable with $\Upsilon(10753)\to\omega \chi_{bJ}$~\cite{Belle-II:2022xdi} observed at Belle II.
Furthermore, the $\epem\to\eta\Upsilon(nS)$ process provides a different approach to search for new bottomonium resonances. 
A measurement of the Born cross section of $\epem\to B^{(*)}\bar B^{(*)}$ with $\Upsilon(10753)$ scan data~\cite{Belle-II:2024niz} indicates that there might be a structure near the $B^{*}\bar B^{*}$ threshold.
In the charmonium sector, a similar structure $\psi(4040)$ was found near the $D^{*}\bar D^{*}$ threshold and confirmed in $\epem\to\eta J/\psi$~\cite{pdg}. 
Analogously, we can search for the possible state near the $B^{*}\bar B^{*}$ threshold in the $\epem\to\eta\Upsilon(nS)$ process.

The charmonium-like state $Y(4230)$~\cite{pdg}, which is produced in $\epem$ annihilation, has been measured to have unexpectedly large branching fractions to $\pipi\psi(nS)$ and $\omega\chi_{cJ}$, and can also decay to the exotic state $X(3872)$ via radiative transition~\cite{BESIII:2019qvy}.
Analogously, it is possible that the $\Upsilon(10753)$ decays to a bottomonium partner of the $X(3872)$~\cite{Liu:2024ets}, $X_b$, via radiative decay.
The authors of Ref.~\cite{Guo:2014sca} note that the $X_b$ state would be likely to decay to $\pipi\chi_{bJ}$, where $\chi_{bJ}$ will predominantly decay to $\gamma\Upsilon(1S)$~\cite{pdg}.

To search for the $\epem\to\eta\Upsilon(nS)$ and $X_b\to\pipi\chi_{bJ}$ events, we reconstruct $\gamma\gamma\pipi\LL,~\ell = e,~\mu$ final states.
For the $\eta\Upsilon(1S)$ channel, the $\eta$ decays to $\pi^+\pi^-\pi^0$, where $\pi^0$ decays to $\gamma\gamma$. In the $\eta\Upsilon(2S)$ channel, the decay modes are either $\eta \rightarrow \gamma\gamma$ or $\eta \rightarrow \pi^+\pi^-\pi^0$, with $\Upsilon(2S) \rightarrow \pi^+\pi^-\Upsilon(1S)$ or $\Upsilon(2S) \rightarrow \ell^+\ell^-$.
We reconstruct $\chi_{bJ}$ via the decay to $\gamma\Upsilon(1S)$.
We always reconstruct $\Upsilon(1S)$ in the $\ell^+\ell^-$ final state.
We exclude the decay chain $\eta\to\gamma\gamma$ with $\Upsilon(nS)\to\LL$ due to the large background from quantum electrodynamics (QED) processes.
The decay chain of $\eta\to\pipipi$ with $\Upsilon(2S)\to\pipi\Upsilon(1S)$ is also excluded due to its low signal sensitivity.
We use data collected with the Belle II detector~\cite{b2tdr,b2tip} in November 2021 at center-of-mass (c.m.) energies of $\sqrt{s}=$ $10.653$, $10.701$, $10.746$, and $10.805$ \gev, which form the $\Upsilon(10753)$ scan data set.

The \belletwo detector operates at the SuperKEKB asymmetric-energy electron-positron collider~\cite{skekb} at KEK. 
The \belletwo detector is a nearly $4\pi$ spectrometer consisting of silicon-based vertexing and drift-chamber tracking systems, Cherenkov-light particle identification detectors, and an electromagnetic calorimeter (ECL), situated within a superconducting solenoid providing a 1.5~T axial magnetic field.
The symmetry axis of these detectors, defined as the $z$ axis, is almost coincident with the direction of the electron beam. 

Monte Carlo (MC) simulated data are used to optimize event selection, determine reconstruction 
efficiencies, extraction of signal-resolution functions, and determine the fit models used to extract the signals.
We simulate $\Upsilon(10753)$ decays with {\sc EvtGen}~\cite{evtGen}, using \texttt{PHOTOS}~\cite{Photos} to simulate final-state radiation.
Initial-state radiation (ISR) at next-to-leading order accuracy in quantum electrodynamics is simulated with {\sc phokhara}~\cite{Phokhara}.
The $e^{+}e^{-}\to\eta\Upsilon(nS)$ processes are simulated in P-wave, while $e^{+}e^{-}\to\gamma X_b$ is simulated uniformly in phase-space. 
Detector simulation is performed with Geant4~\cite{geant4}.
Reconstruction of events from simulation and collision data uses the Belle~II analysis software~\cite{Basf2, Basf2-zenodo}.
An inclusive background sample with four times the luminosity of data includes decays of pairs of $B$ mesons, hadronic processes with light quarks ($u,d,s,c$), and low-multiplicity QED processes, \emph{e.g.}, Bhabha scattering~\cite{babayga1,babayaga2,babayaga3,babayaga4,babayaga5}, $\mu^{+}\mu^{-}(\gamma)$~\cite{babayga1,babayaga2,babayaga3,babayaga4,babayaga5}, ISR-produced hadron pairs~\cite{Phokhara}, and four-track events with at least one lepton pair~\cite{aafh1,aafh2}. These simulated processes  are used to check for contamination from possible backgrounds.


Events are selected online by a hardware trigger that uses drift-chamber and calorimeter information with an efficiency greater than $97\%$ for events containing at least three tracks according to simulation~\cite{b2trg}.
In the offline analysis, tracks reconstructed in the final state are required to originate from the vicinity of the interaction point (within $\SI{2}{\centi\meter}$ along the $z$ axis, and $\SI{0.5}{\centi\meter}$ in the radial direction) to remove beam-related backgrounds and incorrectly reconstructed tracks.
Tracks are required to be within the angular acceptance of the drift chamber, \emph{i.e.}, the polar angle with respect to the $z$ axis should lie within $[17.0,~150.0]^\circ$. 
Tracks with momenta less than 1 GeV/{\it{c}} are considered as pions, those greater than 3 GeV/{\it{c}} are considered as leptons, and those in between are excluded.
To suppress the background from $e/\pi$ mis-identification, we require that at least one pion candidate should have a likelihood ratio $\mathcal{L}_{e}/(\mathcal{L}_{e}+\mathcal{L}_{\pi})$ less than 0.1.
To further separate leptons, the lepton candidates with a ratio of deposited energy in the ECL to track momentum greater than 0.7 are regarded as electrons, whereas those with a ratio less than 0.3 are regarded as muons. 


Photons are reconstructed from the ECL clusters within the angular acceptance of the drift chamber. For photons reconstructed in the barrel and forward end-cap, the deposited energy in the ECL must be greater than 20 MeV. In contrast, for photons reconstructed from the backward end-cap, the energy threshold is set at 22.5 MeV.
The barrel corresponds to a polar angle range of [32.2, 128.7]$^\circ$ while the forward and backward end-caps correspond to [12.4, 31.4]$^\circ$ and [130.7, 155.1]$^\circ$, respectively.
In the reconstruction of $\ups(nS)\to\epem$ events, a bremsstrahlung correction is applied by adding the four-momentum of photons within 0.2 rad around the initial electron direction to the four-momentum of the electron.
To suppress photon conversion background, we require \( \cos(\Delta\theta) < 0.98 \) for the polar angle between the two charged pions.
A four-constraint kinematic fit is performed to the $\gamma\gamma\pipi\LL$ candidates by constraining the four-momenta of the final-state system to the initial $\epem$ c.m.\ system, and to further suppress background, a requirement of $\chi^2<80$ is applied. On average, we find 1.07 candidates per event. 
In events with multiple candidates we select the one with the lowest value of $\chi^2$, which retains the correct candidate with an efficiency of 92.1\%.



In selecting $\epem\to\eta\ups(2S)$ with $\eta\to\pipipi$ and $\Upsilon(2S)\to\LL$, we require  $M(\gamma\gamma)\in[0.115,~0.165]~{\rm GeV}/c^2$, and $M(\LL)\in[10.00,~10.06]~\gevcc$.
For $\eta\to\gamma\gamma$ and $\Upsilon(2S)\to\pipi\Upsilon(1S)$ candidates, we require $M(\LL)\in[9.40,~9.52]~\gevcc$, and $M(\pipi\Upsilon(1S))\in[10.015,~10.030]~\gevcc$, where $M(\pipi\Upsilon(1S))\equiv M(\pipi\LL)-M(\LL)+m(\Upsilon(1S))$ is implemented to improve the mass resolution. 
The regions are chosen to cover approximately 95\% of signal events. 
The $\Upsilon(2S)$ sidebands are defined so that each has the same size as the signal region; the gap between the signal region and the sideband is half the size of the signal region.
To improve the mass resolution, corrected masses $M’(\gamma\gamma)\equiv M(\gamma\gamma)+M(\LL)-m(\Upsilon(1S))$ and $M(\pipipi)\equiv M(\pipi\gamma\gamma)-M(\gamma\gamma)+m(\pi^0)$ are used instead of the invariant masses.
Resolutions are improved by approximately a factor of two, and
the methods for improving the resolution differ between $M'(\gamma\gamma)$ and $M(\pipipi)$ due to their distinct mass construction topologies.
After applying the selection criteria, the two-dimensional distributions of $M(\LL)$ and $M(\pipi\Upsilon(1S))$ versus $M'(\gamma\gamma)$ and $M(\pipipi)$ are shown in Fig.~\ref{fig:2Dplot_etay2s}, summing over all c.m. energies in the $\Upsilon(10753)$ scan data.
Nominal $\Upsilon(2S)$ signal and side-band regions are also displayed in Fig.~\ref{fig:fitTo_etaY2S}.
A clear cluster can be seen in the $\Upsilon(2S)$ signal region.
No peak is observed in the $M'(\gamma\gamma)$ and $M(\pipipi)$ distribution in either the $\Upsilon(2S)$ data sideband or the inclusive background MC sample.

\begin{figure}[ht]
\begin{flushright}
\begin{center}
    \includegraphics[width=0.48\textwidth]{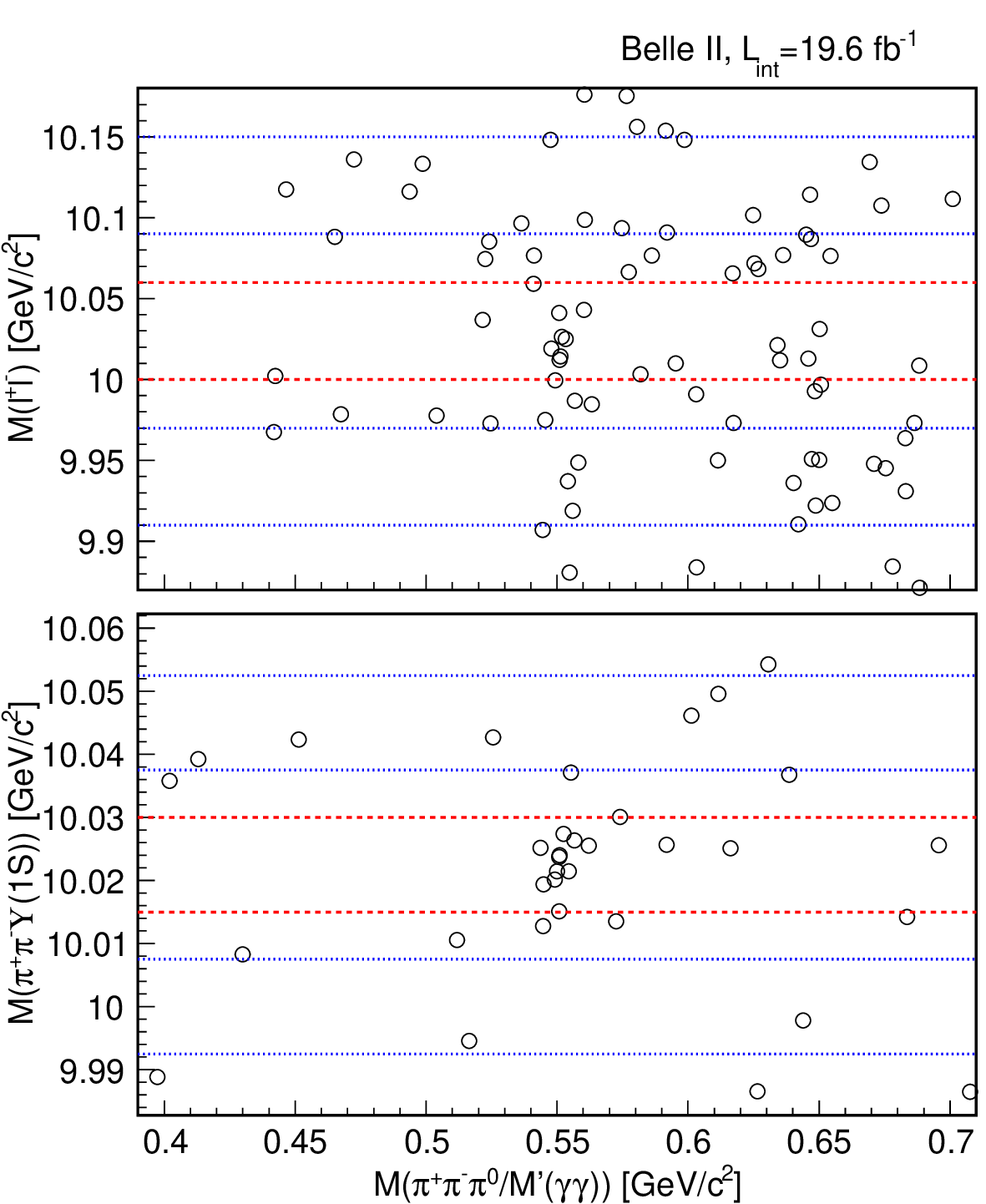}
   
\end{center}
\caption{Two dimensional distribution of $M(\LL)$ versus $M(\pipipi)$ (top) and $M(\pipi\Upsilon(1S))$ versus $M'(\gamma\gamma)$ (bottom) from data. Horizontal lines represents the $\Upsilon(2S)$ signal (red) and sideband boundaries (blue).}\label{fig:2Dplot_etay2s}
\end{flushright}
\end{figure}

\begin{figure*}[ht]
\begin{center}
    \includegraphics[width=0.24\textwidth]{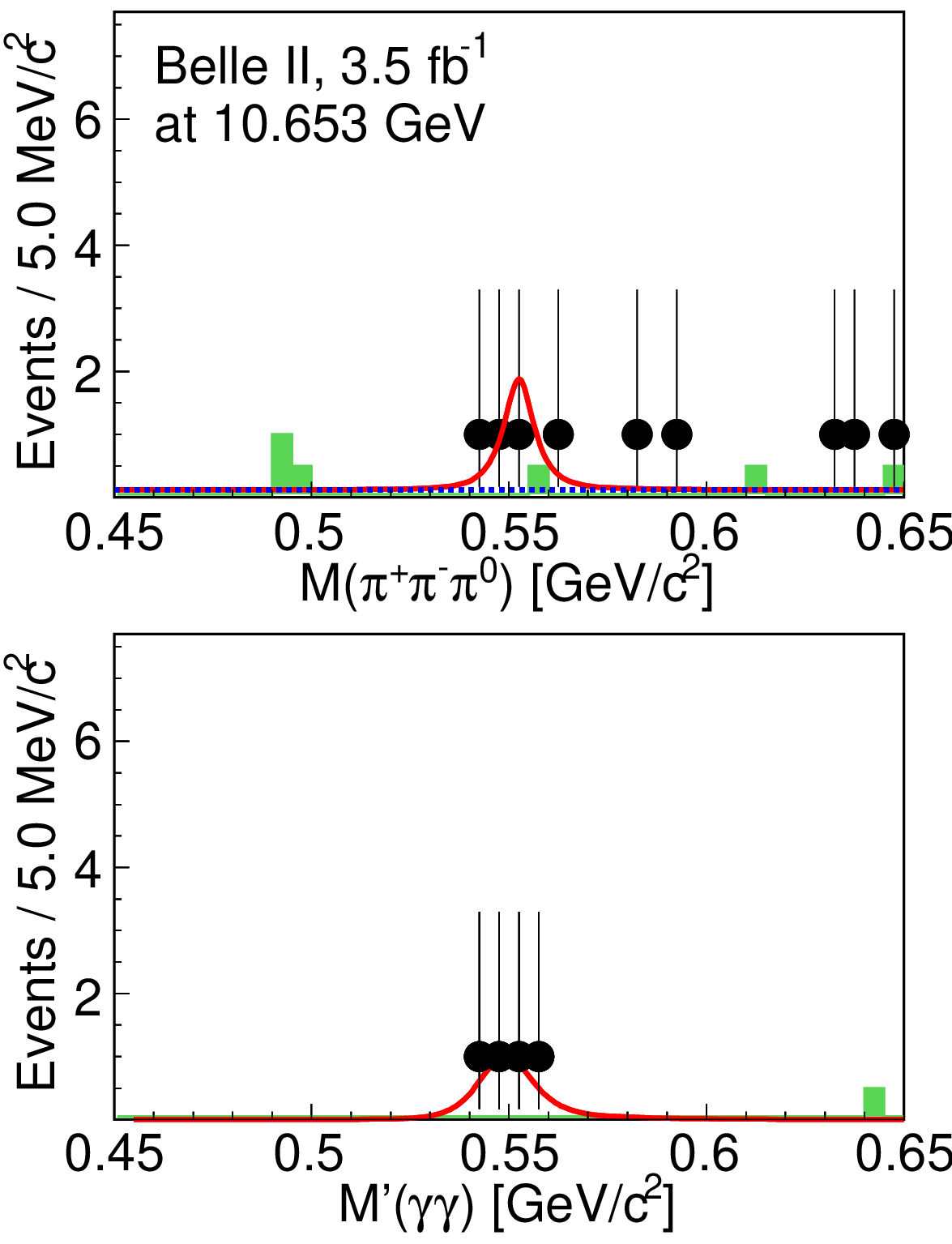}
    \includegraphics[width=0.24\textwidth]{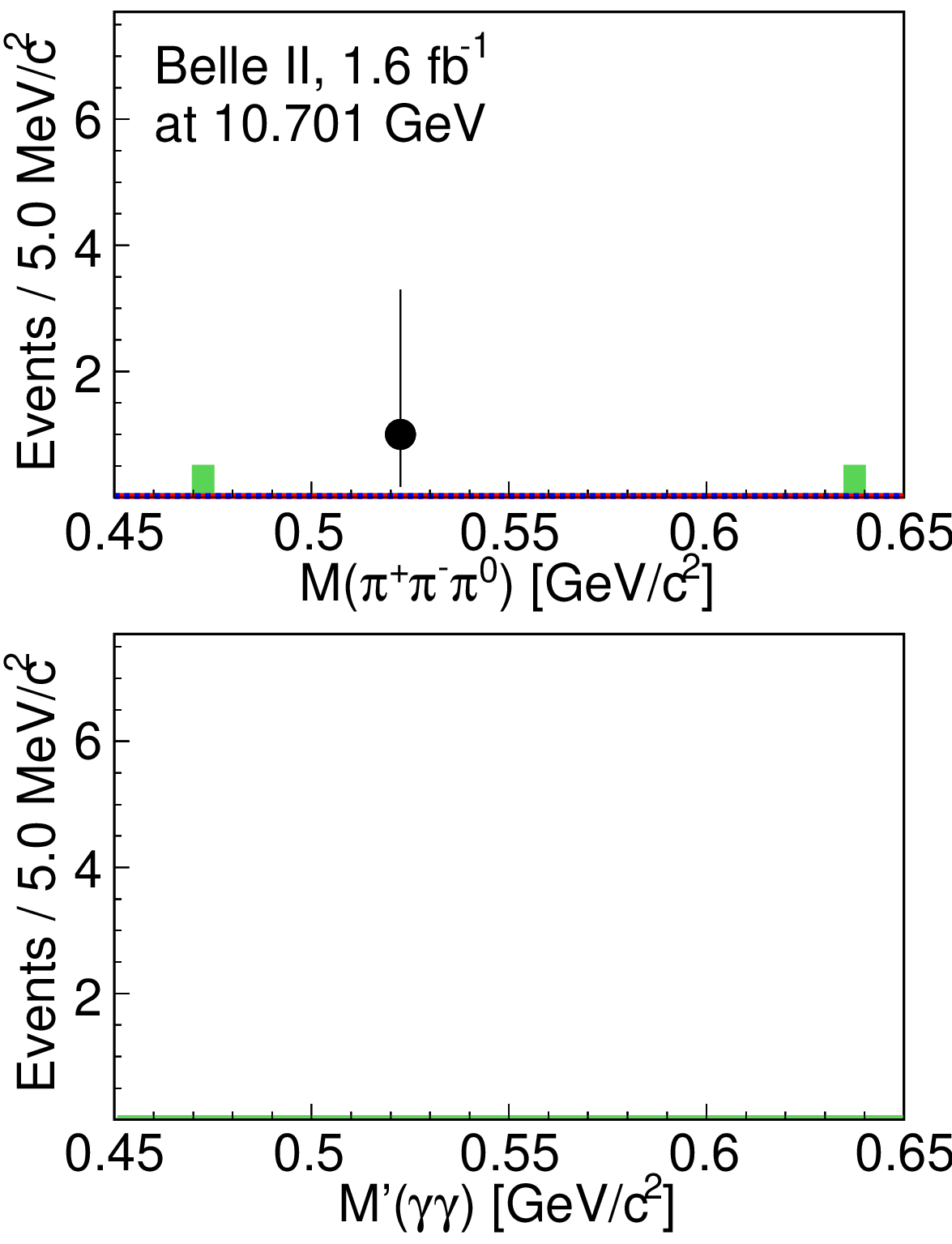}
    \includegraphics[width=0.24\textwidth]{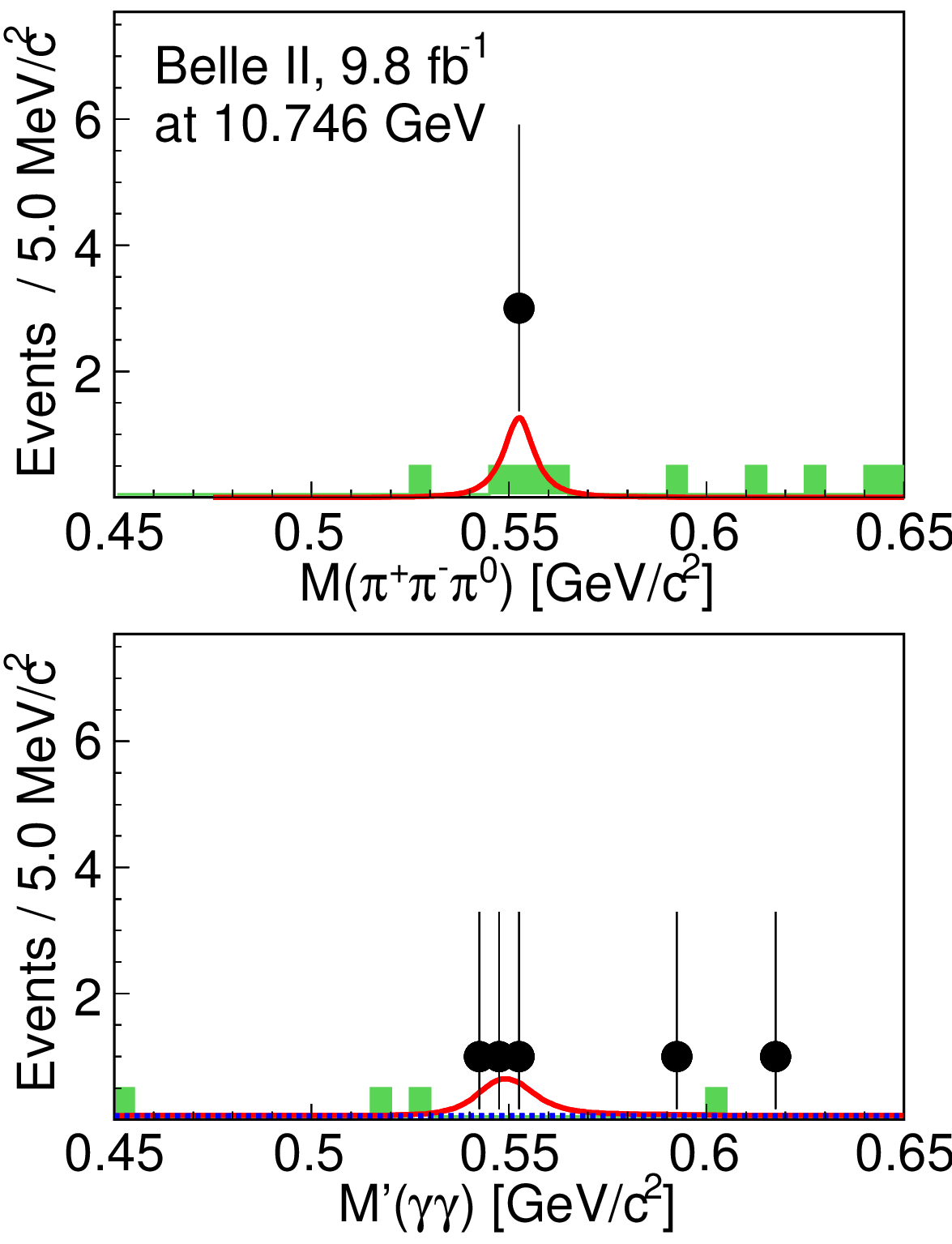}
    \includegraphics[width=0.24\textwidth]{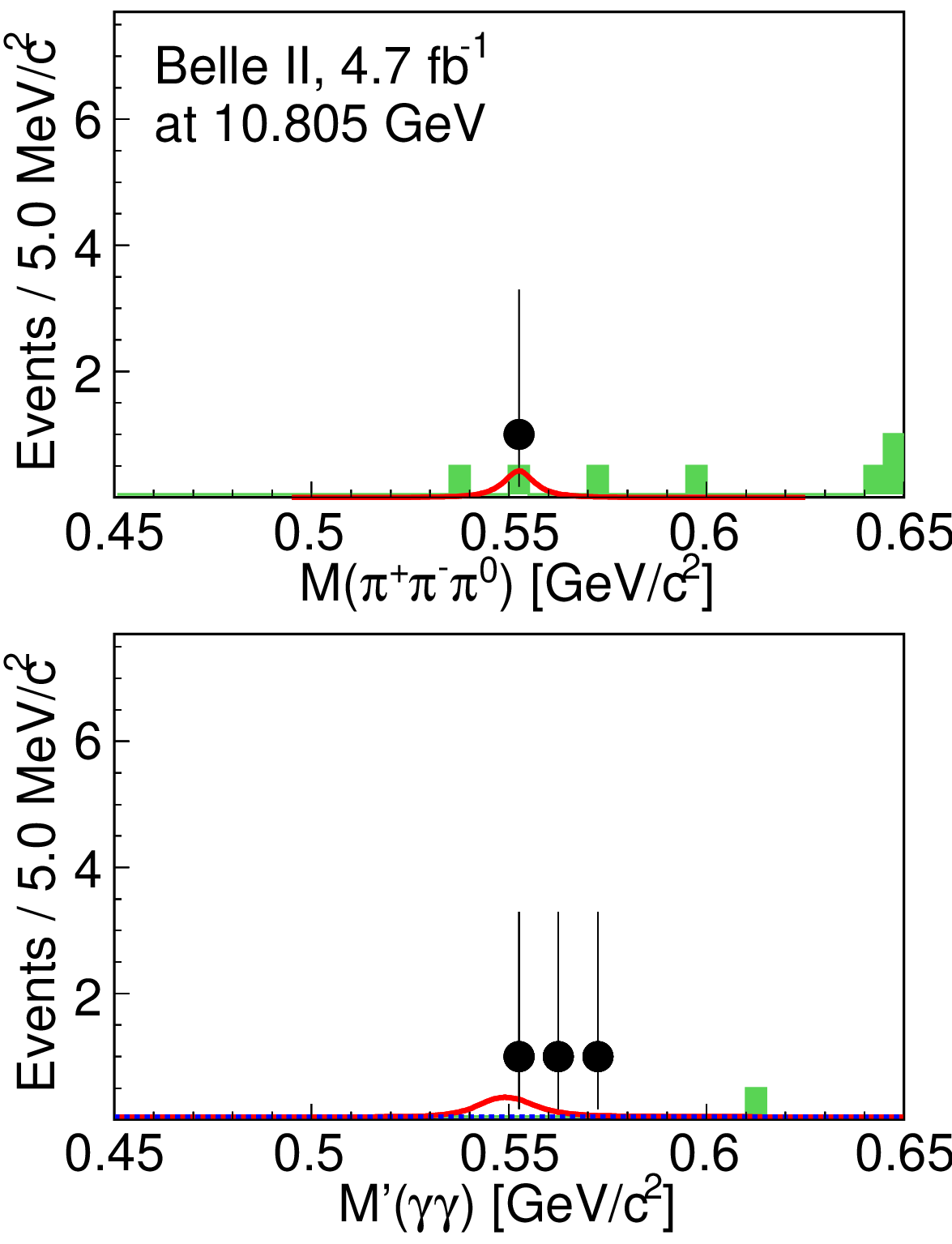}
\end{center}
\caption{Simultaneous fit to $M(\pipipi)$ (top) and $M'(\gamma\gamma)$ (bottom) from each energy point. Plots in the same column correspond to the same scan energy. Dots with error bars are experimental data, the red solid lines represent the fit result, and the green histograms represent events from the $\Upsilon(2S)$ sideband.}\label{fig:fitTo_etaY2S}
\end{figure*}  

The distributions of $M'(\gamma\gamma)$ and $M(\pipipi)$ at the individual energy points are shown in Fig.~\ref{fig:fitTo_etaY2S}.
An extended unbinned maximum likelihood fit is performed simultaneously on $M'(\gamma\gamma)$ and $M(\pipipi)$ at each energy point, 
in which the number of $\epem\to \eta\Upsilon(2S)$ ($N_{\rm prod}$) events is the common free parameter, 
and the relative $\epem\to\eta\Upsilon(2S)$ signal yields ($N_{\rm sig}$) from the two spectra are computed according to the $\eta\to\gamma\gamma$ and $\eta\to\pipipi$ branching fractions ($\mathcal{B}$) and reconstruction efficiencies ($\epsilon$), {\it i.e.}, $N_{\rm sig} = N_{\rm prod} \mathcal{B}\epsilon$.
The background probability density functions (PDFs) are first-order polynomials, and the signal PDFs are the MC histograms.
The fit results are shown in Table~\ref{tab:born_results} as well as
the statistical significance of the signal, which is determined by comparing the change of the likelihood and the number of degrees of freedom in the fit with and without the signal component in the fit. 
Profile likelihood distributions can be found in the Supplemental Material and on HEPData~\cite{HEPdata}.
In a similar fit to the combined data sample, the significance is $6.4\sigma$ (see the Supplemental Material).

The Born cross-section at each energy point is calculated using 
$\sigma_{\rm B} = \frac{N_{\rm prod}\,|1-\Pi|^2} {{\cal L} \,(1+\delta)},$
where ${\cal L}$ is the integrated luminosity of the individual dataset, 
$|1-\Pi|^2$ is the vacuum polarization factor~\cite{VP}, 
and $(1+\delta)$ is the radiative correction factor~\cite{theory_ISR2} calculated using the Born cross-section lineshape in the same way as in Ref.~\cite{belle2_pipiYnS}.
The numerical results of the Born cross-sections are listed in Table~\ref{tab:born_results} and plotted in Fig.~\ref{fig:xsfit}.
Additional analysis with 711 $\rm fb^{-1}$ of data taken around the $\Upsilon(4S)$ resonant peak by Belle~\cite{PBFbook} is performed, 
and the cross section at the $\Upsilon(4S)$ peak is measured to be less than 0.01 pb (with statistical uncertainty only), indicating that the signal events at 10.653 GeV are not coming from the $\Upsilon(4S)$ tail.
We fit the Born cross-sections from both this work and the Belle study of $\Upsilon(5S)\to\eta\Upsilon(2S)$~\cite{Belle:2021gws} as a function of center-of-mass energy $\sqrt{s}$ with three coherent relativistic Breit-Wigner functions representing $\Upsilon(5S)$, $\Upsilon(10753)$, and a possible state near the $B^{*}\bar B^{*}$ threshold.
The mass and width of $\Upsilon(5S)$ and $\Upsilon(10753)$ are fixed to the known values~\cite{pdg,belle2_pipiYnS}.
The mass and width of the new state are taken from the fit to the {$B^{(*)}\bar B^{(*)}$} cross section lineshape reported in Ref.\cite{Belle-II:2024niz}.
The $\chi^2$ of this fit is constructed from the Born cross-section-dependent profile likelihood distribution from the individual simultaneous fit.
We also test hypotheses that include only the $\Upsilon(5S)$ resonance or only the $\Upsilon(5S)$ and $\Upsilon(10753)$ resonances, as shown in Fig.~\ref{fig:xsfit}. These alternative hypotheses are rejected relative to the three-resonance fit at the level of at least 3.6$\sigma$.
Additionally, we investigated an alternative hypothesis in which the contribution from $\Upsilon(10753)$ or the proposed new state is replaced by a flat continuum contribution. Although this hypothesis could not be definitively rejected, it would necessitate a substantial continuum production of $e^{+}e^{-} \to \eta \Upsilon(2S)$ on the same scale as the known decay processes $\Upsilon(10753) \to \pi\pi \Upsilon(nS)$ and $\Upsilon(10753) \to \omega \chi_{b1,2}$~\cite{belle2_pipiYnS, Belle-II:2022xdi}.

\begin{figure}[ht]
\begin{flushleft}
    \includegraphics[width=0.95\linewidth]{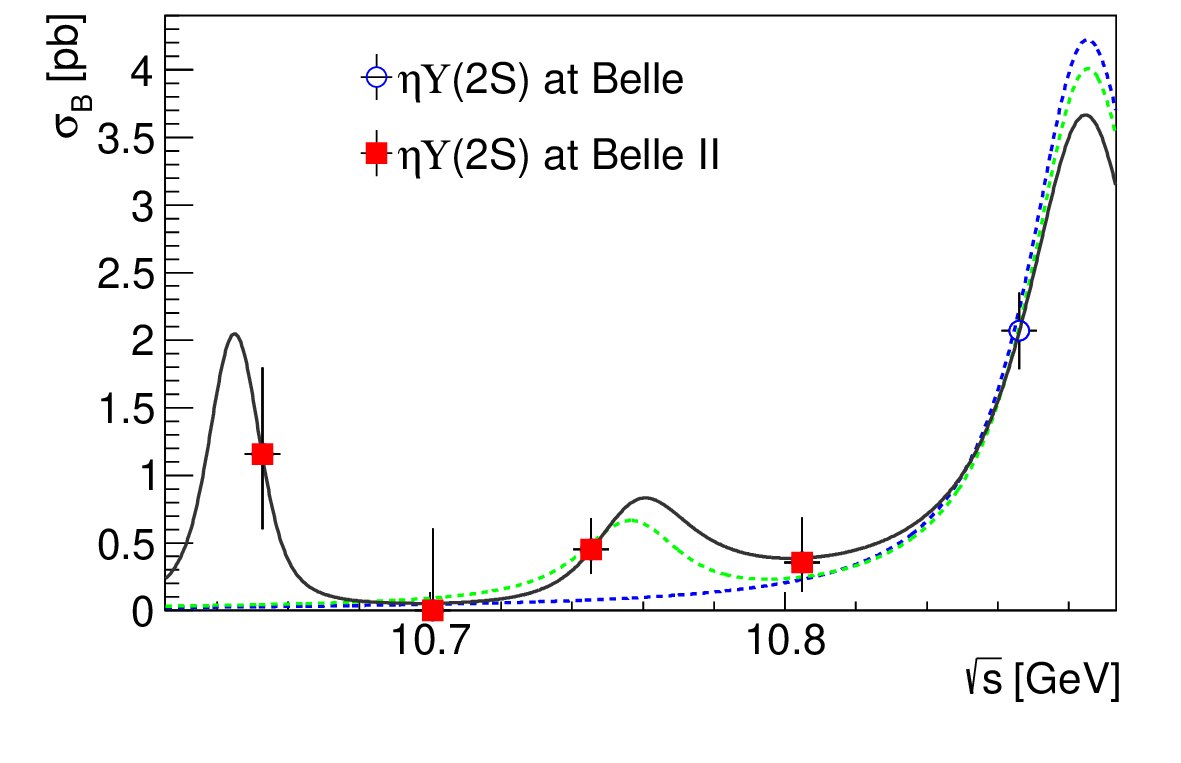} 
\end{flushleft}
  \caption{Born cross-sections for $\epem\to\eta\Upsilon(2S)$ with fit results overlaid. Points with error bars show measured cross sections, the solid curve is the nominal fit result with $\Upsilon(5S)$, $\Upsilon(10753)$ and a state near $B^*\bar B^*$ threshold, the green dashed curve is the fit result with $\Upsilon(5S)$ and $\Upsilon(10753)$, and the blue dashed curve is the fit result with only $\Upsilon(5S)$. The new state near the $B^*\bar B^*$ threshold is decoupled from the other two states, with only its amplitude contributing to the fit.}

  \label{fig:xsfit}

\end{figure}

We use the same $\gamma\gamma\pipi\LL$ final state to reconstruct $\epem\to\eta[\to\pipipi]\Upsilon(1S)$.
To purify the signal, we apply the same selection criteria and mass selection on $\gamma\gamma$ and $\LL$, which are reconstructed as $\pi^0$ and $\Upsilon(1S)$.
Only two events are found in the $\eta$ signal region [0.52, 0.57] GeV$/c^2$ with the full data sample~\cite{SM}. 
Therefore, upper limits at 90\% credibility level (C.L.) of the signal yields at individual energies are estimated using method described in Ref.~\cite{Rolke:2004mj}.
The results for the signal Born cross section, calculated assuming $1/s$ energy dependence, are shown in Table~\ref{tab:born_results}.

 \begin{table}[h]
 	\small
  \caption{Summary of the c.m.\ energy-dependent produced signal yield ($N_{\rm prod}$) and its statistical significance, ISR correction factors ($1+\delta$), weighted efficiencies ($\epsilon$), and Born cross-sections or upper limits at 90\% C.L. ($\sigma^{\rm UL}_{\rm B}$). There are two efficiencies for $\eta\Upsilon(2S)$ channel: the $\eta \to \gamma\gamma$ mode (left) and $\eta \to \pipipi$ mode (right). For $\gamma X_b$ channel, $\sigma_{\rm B}^{(\rm UL)}$ is the upper limit of $\sigma(e^+ e^- \to \gamma X_b) \times \mathcal{B}(X_b \to \pi^+ \pi^- \chi_{bJ})$ at 90\% C.L.
 Uncertainty values for the signal yields are statistical only. For the Born cross sections, the first uncertainty is statistical and the second systematic.}
 	\begin{center}  
 		  \begin{tabular}{l c c c c c }
 			\hline Mode & $N_{\rm prod}$ ($\times10^3$)  &   $(1+\delta)$ & $\epsilon(\%)$ & $\sigma_{\rm B}^{(\rm UL)}$ (pb)\\
 			\hline \multicolumn{3}{l}{$\sqrt{s}=(10653.30\pm1.14)$ MeV,} & \multicolumn{2}{l}{$\mathcal{L}=3.512~\rm fb^{-1}$}  &   \\
 			\hline
 			$\eta\Upsilon(2S)$ & $(3.7^{+1.6}_{-1.3}),4.2\sigma$ &   0.843 & 19.2/15.1 &   $1.16^{+0.51}_{-0.41}\pm0.38$ \\
 			$\eta\Upsilon(1S)$ & $<0.4 $ &  0.895 & 23.9 &   $<0.10$ \\
 			$\gamma X_b$      & $<0.3 $  &  0.784 & 32.0 &   $<0.14$ \\
 			\hline \multicolumn{3}{l}{$\sqrt{s}=(10700.90\pm0.63)$ MeV,} & \multicolumn{2}{l}{$\mathcal{L}=1.632~\rm fb^{-1}$} &   \\
 			\hline
 			$\eta\Upsilon(2S)$ & $(0.0^{+1.0}_{-0.0})$ &        1.691 & 13.2/7.6 &   $0.00^{+0.34}_{-0.00}\pm 0.50$   \\
 			$\eta\Upsilon(1S)$ & $<0.4 $     &     0.901 & 24.0 &   $<0.22$  \\
 			$\gamma X_b$      & $<0.1 $    &    0.803 & 31.3 &   $<0.09$ \\
 			\hline
 			\multicolumn{3}{l}{$\sqrt{s}=(10746.30\pm0.48)$ MeV,} & \multicolumn{2}{l}{$\mathcal{L}=9.818~\rm fb^{-1}$} &   \\
 			\hline
 			$\eta\Upsilon(2S)$ & $(3.2^{+1.6}_{-1.2}),4.8\sigma$  &      0.673 & 17.4/14.2 &   $0.45^{+0.23}_{-0.17}\pm 0.05$  \\
 			$\eta\Upsilon(1S)$ & $<0.9 $  &     0.906 & 23.8 &   $<0.09$ \\
 			$\gamma X_b$      & $<1.4 $    &    0.817 & 29.8 &   $<0.17$ \\
 			\hline
 			\multicolumn{3}{l}{$\sqrt{s}=(10804.50\pm0.70)$ MeV,} & \multicolumn{2}{l}{$\mathcal{L}=4.689~\rm fb^{-1}$}  &   \\
 			\hline
 			$\eta\Upsilon(2S)$ & $(1.5^{+1.3}_{-0.9}),2.8\sigma$ &           0.822 & 17.1/15.2 &   $0.36^{+0.33}_{-0.21}\pm 0.04$ \\
 			$\eta\Upsilon(1S)$ & $<0.4 $  &      0.912 & 24.6 &   $<0.08$ \\
 			$\gamma X_b$      & $<1.3 $  &    0.833 & 28.2 &   $<0.32$ \\
 			\hline
 		\end{tabular}
 	\end{center}
 	\label{tab:born_results}
 \end{table}  

We also search for the $X_b$ with $X_b\to\pi\pi\chi_{bJ}$ in $\epem\to\gamma\X_{b}$ with the same final state.
We do not separate $\chi_{b1}$ and $\chi_{b2}$ because of their
very close invariant masses.
The search is performed in the mass window $M(\gamma\Upsilon(1S))\in[9.84,~9.96]~\gevcc$.
The signal PDFs are the MC histograms. The dominant backgrounds for these channels are $\epem\to\eta\Upsilon(1,2S)$, $\epem\to\omega[\to\pipipi]\chi_{b1,2}$, and $\epem\to\pipi\Upsilon(2S)[\to\gamma\chi_{bJ}]$, where subsequent decay final states are shown in square brackets.
The cross section of $\epem\to\eta\Upsilon(1,2S)$ is reported in Table~\ref{tab:born_results} and the contribution to $\pi\pi\chi_{bJ}$ is small.
The other two backgrounds are difficult to suppress because of the similarity in phase-space, thus their contributions are constrained during the fit according to the measured Born cross-sections~\cite{Belle-II:2022xdi,belle2_pipiYnS}, branching fractions into the final state $\gamma\gamma\pipi\LL$~\cite{pdg}, and reconstruction efficiencies obtained from the MC study.
An unbinned maximum likelihood fit is performed to the invariant mass $M(\pipi\chi_{b1})$~\cite{SM}.
Different hypotheses of the mass of $X_b$ are evaluated, and the maximum probability is found when $m(X_b)=10.50~\rm GeV/c^2$. No evident signal of $X_b$ is found.
The upper limits of $\sigma(e^+ e^- \to \gamma X_b) \times \mathcal{B}(X_b \to \pi^+ \pi^- \chi_{bJ})$ at 90\% C.L. are estimated using a Bayesian method as described in Ref.~\cite{belle2_pipiYnS} with the assumption that $M(X_b)=10.50~{\rm GeV}/c^2$, as shown in Table~\ref{tab:born_results}.


Sources of systematic uncertainty for the cross-section measurements include tracking and photon detection efficiencies, the integrated luminosity of the sample ($\mathcal{L}$), the choice of simulated-event generators, ISR factors, the chosen generator decay model, trigger efficiency, the cascade branching fractions ($\mathcal{B}$), and the fit procedure.

In the simulation of $\epem\to\gamma X_{b}$, we change the uniformly-distributed radiative transition to alternative models.
The change of detector efficiency due to variation of the decay model is taken as the systematic uncertainty. 
This uncertainty is not considered for the process $\epem\to\eta\Upsilon(nS)$, which is known to have a P-wave distribution.
For the ISR factors in $\epem\to\eta\Upsilon(2S)$,
we take the changes of the Born cross-sections in different lineshape hypotheses as the systematic uncertainties.
We note that the systematic uncertainty is up to 33.8\% for \( e^+e^- \to \eta \Upsilon(2S) \) at 10.653 GeV, arising mainly from the presence or absence of a new state near the \( B^*\bar{B}^* \) threshold.

To evaluate the systematic uncertainty due to fit procedure, we expand the fit ranges and change the background parameterization from a linear function to a uniform or a quadratic function, or the parameterized shape of the $\Upsilon(2S)$ sideband, taking the differences in the signal yields between the nominal fit and the alternative fit as the systematic uncertainty.
We find that the significance of the observation of $\epem\to\eta\Upsilon(2S)$ is always greater than $6.0\sigma$ in all fit ranges and background parameterizations. 
For $\epem\to\gamma X_b$, to be conservative, the largest upper limits are taken as the nominal results from fits with various fitting ranges.

From studies of $\epem\to\mumu$
and $e^{+}e^{-}\to \pi^{+}\pi^{-}\pi^{+}\pi^{-}\pi^{0}\pi^{0}$, 
we assign a systematic uncertainty of $1\%$ to the trigger efficiency simulation. 
Tracking uncertainty of 0.3\% per track is obtained from $\bar B^0 \to D^{*+}$[$\to D^0 \pi^+$]$ \pi^{-}$ and $e^+e^-\to \tau^+\tau^-$ control samples in data. 
The uncertainty on the photon detection efficiency, studied with $\epem\to\mu^+\mu^-\gamma_{\rm ISR}$, is 3.5\% per photon.
The uncertainty on the reconstruction efficiency for the $\pi^{0}$ produced in $\eta\to\pipipi$ is 4.3\% based on studies of $B^+\to\pi^+ \bar D^{*0},~\bar D^{*0}\to \bar D^0 \pi^0$.
The uncertainty of the integrated luminosity is 0.7\%, and the uncertainty from the choice of generator is estimated to be 1.5\%~\cite{theory_ISR}.
The uncertainties on pion and electron identification efficiencies are negligible compared to other systematic uncertainties.
The uncertainties in all the branching fractions used in this analysis are taken from Ref.~\cite{pdg}. 
All systematic uncertainties are combined in quadrature~\cite{SM}.
The Born cross sections or their upper limits after considering the systematic uncertainties are summarized in Table~\ref{tab:born_results}.

In summary, we observe the process $\epem\to\eta\Upsilon(2S)$ using four samples of $\epem$ collision data taken at c.m. energies near the $\Upsilon(10753)$ with a significance greater than $6.0\sigma$.
No significant signal for $\Upsilon(10753)$ is observed in the fit to the Born cross sections.
Hypotheses that attribute the full signal to the $\Upsilon(5S)$ and/or the $\Upsilon(10753)$ states are rejected at the level of at least 3.6$\sigma$. 
This favors the presence of a new state near 10.65 GeV or substantial continuum production, motivating further data collection in this energy region. 
The new state could be a $B^*\bar{B}^*$ molecule suggested in Ref.~\cite{Belle-II:2024niz} to explain the rapid rise of the $\epem\to B^*\bar{B}^*$ cross section above the corresponding threshold. 
The production ratio of $\epem\to\eta\Upsilon(2S)$ to $\epem \to  \pipi\Upsilon(2S)$ near the $\Upsilon(10753)$ peak at $\sqrt{s}=10.746~\rm GeV$ is $0.31^{+0.17}_{-0.12} \pm 0.03$, and near the $B^*\bar{B}^*$ threshold at $\sqrt{s}=10.653~\rm GeV$ we determine the lower limit $>22.6$ at 90\% C.L.  These imply a strong violation of heavy quark spin symmetry~\cite{Bondar:2016hva}, which predicts a production ratio of order $10^{-3}$, and might be a further indication of a $B^*\bar{B}^*$ molecular state.

We also search for the process $\epem\to\eta\Upsilon(1S)$, as well as for $\epem\to\gamma X_{b}$, where $X_b$ is the bottomonium-sector partner of the $X(3872)$.
No evidence of a $\epem\to\eta\Upsilon(1S)$ or $\epem\to\gamma X_b$ signal is found.
The upper limits of $\epem\to\eta\Upsilon(1S)$ and $\gamma X_b$ with $X_b\to\pipi\chi_{bJ, J=1,2}$ are estimated at 90\% C.L.

This work, based on data collected using the Belle II detector, which was built and commissioned prior to March 2019,
was supported by
Higher Education and Science Committee of the Republic of Armenia Grant No.~23LCG-1C011;
Australian Research Council and Research Grants
No.~DP200101792, 
No.~DP210101900, 
No.~DP210102831, 
No.~DE220100462, 
No.~LE210100098, 
and
No.~LE230100085; 
Austrian Federal Ministry of Education, Science and Research,
Austrian Science Fund (FWF) Grants
DOI:~10.55776/P34529,
DOI:~10.55776/J4731,
DOI:~10.55776/J4625,
DOI:~10.55776/M3153,
and
DOI:~10.55776/PAT1836324,
and
Horizon 2020 ERC Starting Grant No.~947006 ``InterLeptons'';
Natural Sciences and Engineering Research Council of Canada, Digital Research Alliance of Canada, and Canada Foundation for Innovation;
National Key R\&D Program of China under Contract No.~2024YFA1610503,
and
No.~2024YFA1610504
National Natural Science Foundation of China and Research Grants
No.~11575017,
No.~11761141009,
No.~11705209,
No.~11975076,
No.~12135005,
No.~12150004,
No.~12161141008,
No.~12405099,
No.~12475093,
and
No.~12175041,
and Shandong Provincial Natural Science Foundation Project~ZR2022JQ02;
the Czech Science Foundation Grant No. 22-18469S,  Regional funds of EU/MEYS: OPJAK
FORTE CZ.02.01.01/00/22\_008/0004632 
and
Charles University Grant Agency project No. 246122;
European Research Council, Seventh Framework PIEF-GA-2013-622527,
Horizon 2020 ERC-Advanced Grants No.~267104 and No.~884719,
Horizon 2020 ERC-Consolidator Grant No.~819127,
Horizon 2020 Marie Sklodowska-Curie Grant Agreement No.~700525 ``NIOBE''
and
No.~101026516,
and
Horizon 2020 Marie Sklodowska-Curie RISE project JENNIFER2 Grant Agreement No.~822070 (European grants);
L'Institut National de Physique Nucl\'{e}aire et de Physique des Particules (IN2P3) du CNRS
and
L'Agence Nationale de la Recherche (ANR) under Grant No.~ANR-21-CE31-0009 (France);
BMFTR, DFG, HGF, MPG, and AvH Foundation (Germany);
Department of Atomic Energy under Project Identification No.~RTI 4002,
Department of Science and Technology,
and
UPES SEED funding programs
No.~UPES/R\&D-SEED-INFRA/17052023/01 and
No.~UPES/R\&D-SOE/20062022/06 (India);
Israel Science Foundation Grant No.~2476/17,
U.S.-Israel Binational Science Foundation Grant No.~2016113, and
Israel Ministry of Science Grant No.~3-16543;
Istituto Nazionale di Fisica Nucleare and the Research Grants BELLE2,
and
the ICSC – Centro Nazionale di Ricerca in High Performance Computing, Big Data and Quantum Computing, funded by European Union – NextGenerationEU;
Japan Society for the Promotion of Science, Grant-in-Aid for Scientific Research Grants
No.~16H03968,
No.~16H03993,
No.~16H06492,
No.~16K05323,
No.~17H01133,
No.~17H05405,
No.~18K03621,
No.~18H03710,
No.~18H05226,
No.~19H00682, 
No.~20H05850,
No.~20H05858,
No.~22H00144,
No.~22K14056,
No.~22K21347,
No.~23H05433,
No.~26220706,
and
No.~26400255,
and
the Ministry of Education, Culture, Sports, Science, and Technology (MEXT) of Japan;  
National Research Foundation (NRF) of Korea Grants
No.~2021R1-F1A-1064008, 
No.~2022R1-A2C-1003993,
No.~2022R1-A2C-1092335,
No.~RS-2016-NR017151,
No.~RS-2018-NR031074,
No.~RS-2021-NR060129,
No.~RS-2022-NR068913,
No.~RS-2023-00208693,
No.~RS-2024-00354342
and
No.~RS-2025-02219521,
Radiation Science Research Institute,
Foreign Large-Size Research Facility Application Supporting project,
the Global Science Experimental Data Hub Center, the Korea Institute of Science and
Technology Information (K25L2M2C3 ) 
and
KREONET/GLORIAD;
Universiti Malaya RU grant, Akademi Sains Malaysia, and Ministry of Education Malaysia;
Frontiers of Science Program Contracts
No.~FOINS-296,
No.~CB-221329,
No.~CB-236394,
No.~CB-254409,
and
No.~CB-180023, and SEP-CINVESTAV Research Grant No.~237 (Mexico);
the Polish Ministry of Science and Higher Education and the National Science Center;
the Ministry of Science and Higher Education of the Russian Federation
and
the HSE University Basic Research Program, Moscow;
University of Tabuk Research Grants
No.~S-0256-1438 and No.~S-0280-1439 (Saudi Arabia), and
Researchers Supporting Project number (RSPD2025R873), King Saud University, Riyadh,
Saudi Arabia;
Slovenian Research Agency and Research Grants
No.~J1-50010
and
No.~P1-0135;
Ikerbasque, Basque Foundation for Science,
State Agency for Research of the Spanish Ministry of Science and Innovation through Grant No. PID2022-136510NB-C33, Spain,
Agencia Estatal de Investigacion, Spain
Grant No.~RYC2020-029875-I
and
Generalitat Valenciana, Spain
Grant No.~CIDEGENT/2018/020;
The Knut and Alice Wallenberg Foundation (Sweden), Contracts No.~2021.0174 and No.~2021.0299;
National Science and Technology Council,
and
Ministry of Education (Taiwan);
Thailand Center of Excellence in Physics;
TUBITAK ULAKBIM (Turkey);
National Research Foundation of Ukraine, Project No.~2020.02/0257,
and
Ministry of Education and Science of Ukraine;
the U.S. National Science Foundation and Research Grants
No.~PHY-1913789 
and
No.~PHY-2111604, 
and the U.S. Department of Energy and Research Awards
No.~DE-AC06-76RLO1830, 
No.~DE-SC0007983, 
No.~DE-SC0009824, 
No.~DE-SC0009973, 
No.~DE-SC0010007, 
No.~DE-SC0010073, 
No.~DE-SC0010118, 
No.~DE-SC0010504, 
No.~DE-SC0011784, 
No.~DE-SC0012704, 
No.~DE-SC0019230, 
No.~DE-SC0021274, 
No.~DE-SC0021616, 
No.~DE-SC0022350, 
No.~DE-SC0023470; 
and
the Vietnam Academy of Science and Technology (VAST) under Grants
No.~NVCC.05.02/25-25
and
No.~DL0000.05/26-27.

These acknowledgements are not to be interpreted as an endorsement of any statement made
by any of our institutes, funding agencies, governments, or their representatives.

We thank the SuperKEKB team for delivering high-luminosity collisions;
the KEK cryogenics group for the efficient operation of the detector solenoid magnet and IBBelle on site;
the KEK Computer Research Center for on-site computing support; the NII for SINET6 network support;
and the raw-data centers hosted by BNL, DESY, GridKa, IN2P3, INFN, 
and the University of Victoria.


\begin{thebibliography}{**}

\bibitem{pdg}
S.~Navas \textit{et al.} [Particle Data Group],
Phys. Rev. D \textbf{110}, 030001 (2024).

\bibitem{exp_belle}
R.~Mizuk \textit{et al.} [Belle],
JHEP \textbf{10}, 220 (2019)
[arXiv:1905.05521 [hep-ex]].

\bibitem{Belle-II:2022xdi}
I.~Adachi \textit{et al.} [Belle-II],
Phys. Rev. Lett. \textbf{130}, 091902 (2023)
[arXiv:2208.13189 [hep-ex]].

\bibitem{belle2_pipiYnS}
I.~Adachi \textit{et al.} [Belle-II],
JHEP \textbf{07}, 116 (2024)
[arXiv:2401.12021 [hep-ex]].

\bibitem{Belle-II:2023twj}
I.~Adachi \textit{et al.} [Belle-II],
Phys. Rev. D \textbf{109}, 072013 (2024)
[arXiv:2312.13043 [hep-ex]].

\bibitem{Bai:2022cfz}
Z.~Y.~Bai, Y.~S.~Li, Q.~Huang, X.~Liu and T.~Matsuki,
Phys. Rev. D \textbf{105}, no.7, 074007 (2022)
[arXiv:2201.12715 [hep-ph]].

\bibitem{Li:2021jjt}
Y.~S.~Li, Z.~Y.~Bai, Q.~Huang and X.~Liu,
Phys. Rev. D \textbf{104}, 034036 (2021)
[arXiv:2106.14123 [hep-ph]].

\bibitem{BESIII:2019qvy}
M.~Ablikim \textit{et al.} [BESIII],
Phys. Rev. Lett. \textbf{122}, 232002 (2019)
[arXiv:1903.04695 [hep-ex]].

\bibitem{Liu:2024ets}
S.~D.~Liu, H.~D.~Cai, Z.~X.~Cai, H.~S.~Gao, G.~Li, F.~Wang and J.~J.~Xie,
Phys. Rev. D \textbf{109}, 094045 (2024)
[arXiv:2403.01676 [hep-ph]].

\bibitem{Guo:2014sca}
F.~K.~Guo, U.~G.~Mei{\ss}ner, W.~Wang and Z.~Yang,
Eur. Phys. J. C \textbf{74}, 3063 (2014)
[arXiv:1402.6236 [hep-ph]].


\bibitem{b2tdr}
T.~Abe \textit{et al.} [Belle-II],
arXiv:1011.0352 [physics.ins-det].

\bibitem{b2tip}
E.~Kou \textit{et al.} [Belle-II],
PTEP \textbf{2019}, 123C01 (2019)
[erratum: PTEP \textbf{2020}, 029201 (2020)]
[arXiv:1808.10567 [hep-ex]].

\bibitem{skekb}
K.~Akai \textit{et al.} [SuperKEKB],
Nucl. Instrum. Meth. A \textbf{907}, 188-199 (2018)
[arXiv:1809.01958 [physics.acc-ph]].

\bibitem{evtGen}
D.~J.~Lange,
Nucl. Instrum. Meth. A \textbf{462}, 152-155 (2001).

\bibitem{Photos}
E.~Barberio, B.~van Eijk and Z.~Was,
Comput. Phys. Commun. \textbf{66}, 115-128 (1991).

\bibitem{Phokhara}
G.~Rodrigo, H.~Czyz, J.~H.~Kuhn and M.~Szopa,
Eur. Phys. J. C \textbf{24}, 71-82 (2002)
[arXiv:hep-ph/0112184 [hep-ph]].

\bibitem{geant4}
S.~Agostinelli \textit{et al.} [GEANT4],
Nucl. Instrum. Meth. A \textbf{506}, 250-303 (2003).

\bibitem{Basf2}
T.~Kuhr \textit{et al.} [Belle-II Framework Software Group],
Comput. Softw. Big Sci. \textbf{3}, 1 (2019)
[arXiv:1809.04299 [physics.comp-ph]].

\bibitem{Basf2-zenodo}
The Belle II Collaboration. (2025). Belle II Analysis Software Framework (basf2) (release-09-00-00). Zenodo. \url{https://doi.org/10.5281/zenodo.16268234}.

\bibitem{babayga1}
G.~Balossini, C.~M.~Carloni Calame, G.~Montagna, O.~Nicrosini and F.~Piccinini,
Nucl. Phys. B \textbf{758}, 227-253 (2006)
[arXiv:hep-ph/0607181 [hep-ph]].

\bibitem{babayaga2}
G.~Balossini, C.~Bignamini, C.~M.~C.~Calame, G.~Montagna, O.~Nicrosini and F.~Piccinini,
Phys. Lett. B \textbf{663}, 209-213 (2008)
[arXiv:0801.3360 [hep-ph]].

\bibitem{babayaga3}
C.~M.~Carloni Calame, G.~Montagna, O.~Nicrosini and F.~Piccinini,
Nucl. Phys. B Proc. Suppl. \textbf{131}, 48-55 (2004)
[arXiv:hep-ph/0312014 [hep-ph]].

\bibitem{babayaga4}
C.~M.~Carloni Calame,
Phys. Lett. B \textbf{520}, 16-24 (2001)
[arXiv:hep-ph/0103117 [hep-ph]].

\bibitem{babayaga5}
C.~M.~Carloni Calame, C.~Lunardini, G.~Montagna, O.~Nicrosini and F.~Piccinini,
Nucl. Phys. B \textbf{584}, 459-479 (2000)
[arXiv:hep-ph/0003268 [hep-ph]].

\bibitem{aafh1}
F.~A.~Berends, P.~H.~Daverveldt and R.~Kleiss,
Nucl. Phys. B \textbf{253}, 441-463 (1985).

\bibitem{aafh2}
F.~A.~Berends, P.~H.~Daverveldt and R.~Kleiss,
Comput. Phys. Commun. \textbf{40}, 285-307 (1986).

\bibitem{b2trg}
Y.~Iwasaki, B.~Cheon, E.~Won, X.~Gao, L.~Macchiarulo, K.~Nishimura and G.~Varner,
IEEE Trans. Nucl. Sci. \textbf{58}, 1807-1815 (2011).

\bibitem{HEPdata}
{\textcolor{red}{To follow after journal acceptance.
}}

\bibitem{VP}
S.~Actis \textit{et al.} [Working Group on Radiative Corrections and Monte Carlo Generators for Low Energies],
Eur. Phys. J. C \textbf{66}, 585-686 (2010)
[arXiv:0912.0749 [hep-ph]].

\bibitem{theory_ISR2}
M.~Benayoun, S.~I.~Eidelman, V.~N.~Ivanchenko and Z.~K.~Silagadze,
Mod. Phys. Lett. A \textbf{14}, 2605-2614 (1999)
[arXiv:hep-ph/9910523 [hep-ph]].

\bibitem{PBFbook}
Ed. A. J. Bevan, B. Golob, T. Mannel, S. Prell, and B. D. Yabsley,
Eur. Phys. J. C \textbf{74}, 3026 (2014)
[arXiv:1406.6311 [hep-ex]].

\bibitem{Belle:2021gws}
E.~Kovalenko \textit{et al.} [Belle],
Phys. Rev. D \textbf{104}, no.11, 112006 (2021)
[arXiv:2108.04426 [hep-ex]].

\bibitem{Belle-II:2024niz}
I.~Adachi \textit{et al.} [Belle-II],
JHEP \textbf{10}, 114 (2024)
[arXiv:2405.18928 [hep-ex]].

\bibitem{SM}
See Supplemental Material for the profiled likelihood distributions, signal extraction of $e^+e^-\to\eta\Upsilon(1S)$ and $e^+e^-\to\gamma X_b$, and the systematic uncertainties of Born cross-sections.

\bibitem{Rolke:2004mj}
W.~A.~Rolke, A.~M.~Lopez and J.~Conrad,
Nucl. Instrum. Meth. A \textbf{551}, 493-503 (2005)
[arXiv:physics/0403059 [physics]].

\bibitem{theory_ISR}
E.~A.~Kuraev and V.~S.~Fadin,
Sov. J. Nucl. Phys. \textbf{41}, 466-472 (1985)

\bibitem{Bondar:2016hva}
A.~E.~Bondar, R.~V.~Mizuk and M.~B.~Voloshin,
Mod. Phys. Lett. A \textbf{32} (2017) no.04, 1750025
doi:10.1142/S0217732317500250
[arXiv:1610.01102 [hep-ph]].



\end{thebibliography}

\clearpage

\title{Supplemental Material}

\maketitle

\section{Projection of the $\eta$ distribution mass distribution for the combined data sample}

Projections of invariant masses $M'(\gamma \gamma)$ and $M(\pi^+
\pi^- \pi^0)$ are shown in Fig.~1, with data from all four energy points combined. 
The spectra are fitted with the same method as the individual fit.


\begin{figure*}[ht]
\begin{center}
    \includegraphics[width=0.750\textwidth]{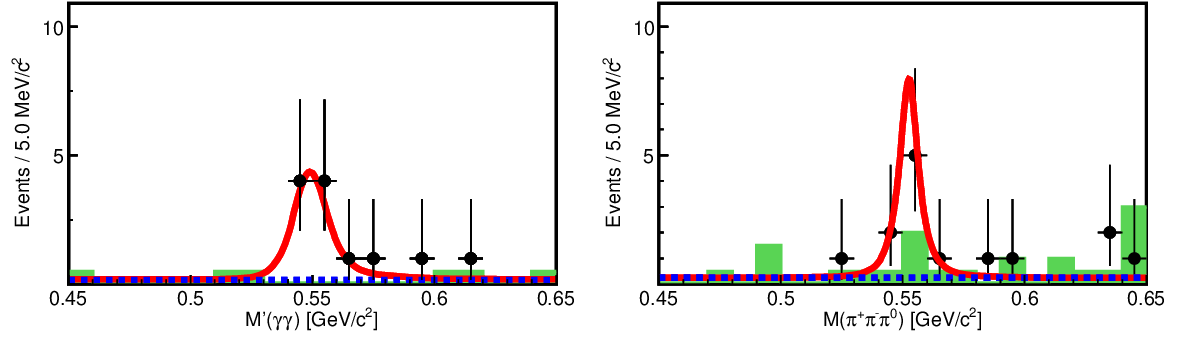} \\
\end{center}
\caption{Simultaneous fit to $M(\gamma\gamma)$ and $M(\pi^+\pi^-\pi^0)$ in the data sample where all four energy points are combined; the green histogram shows events from the $\Upsilon(2S)$ sidebands.}\label{fig:fitTo_etaY2S}
\end{figure*}  

\section{Profile likelihood distributions in $e^+e^-\to\eta\Upsilon(2S)$}

We provide the profiled likelihood distributions as a function of the Born cross sections at each energy point.  
The likelihoods are obtained from the simultaneous fit to  $M'(\gamma\gamma)$ and $M(\pi^+\pi^-\pi^0)$ for the $e^+e^-\to\eta\Upsilon(2S)$ yield.  
These profiled likelihoods are used as input in the Born cross section fit.

\begin{figure*}[h]
\begin{center}
    \includegraphics[width=0.24\textwidth]{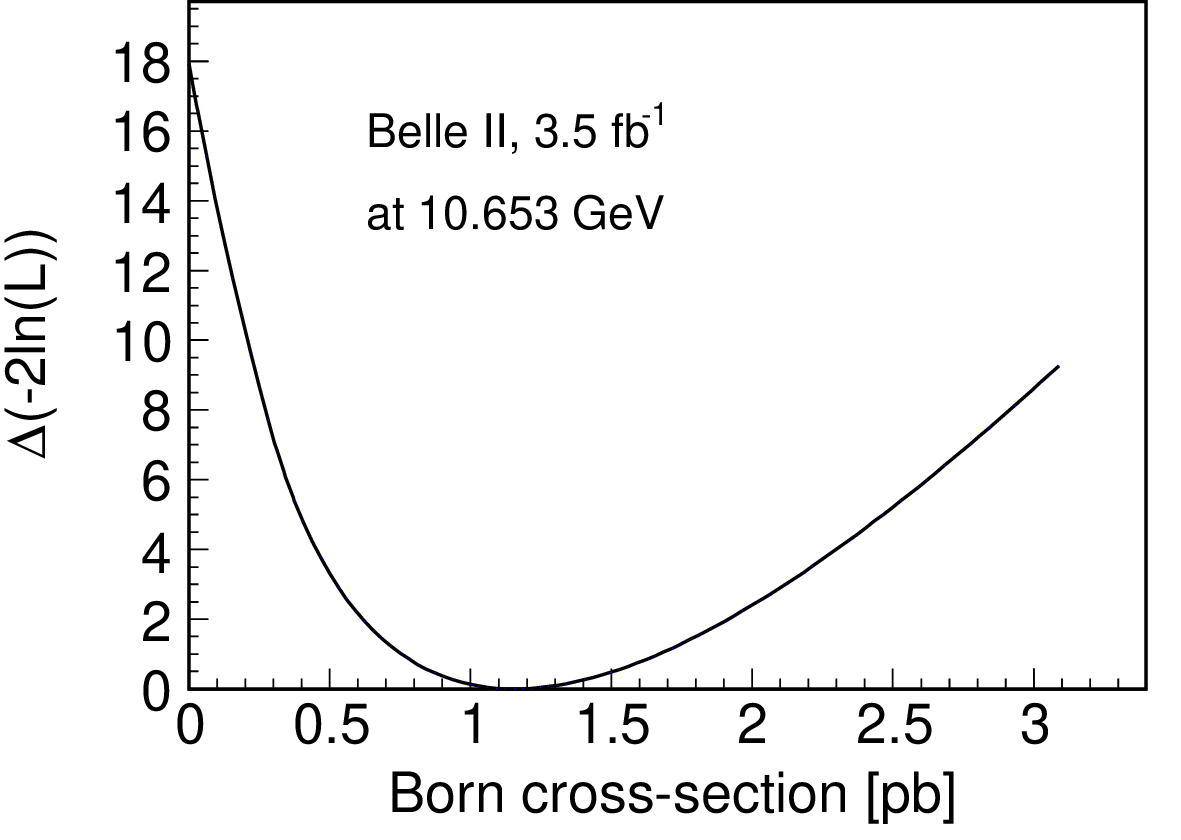}
    \includegraphics[width=0.24\textwidth]{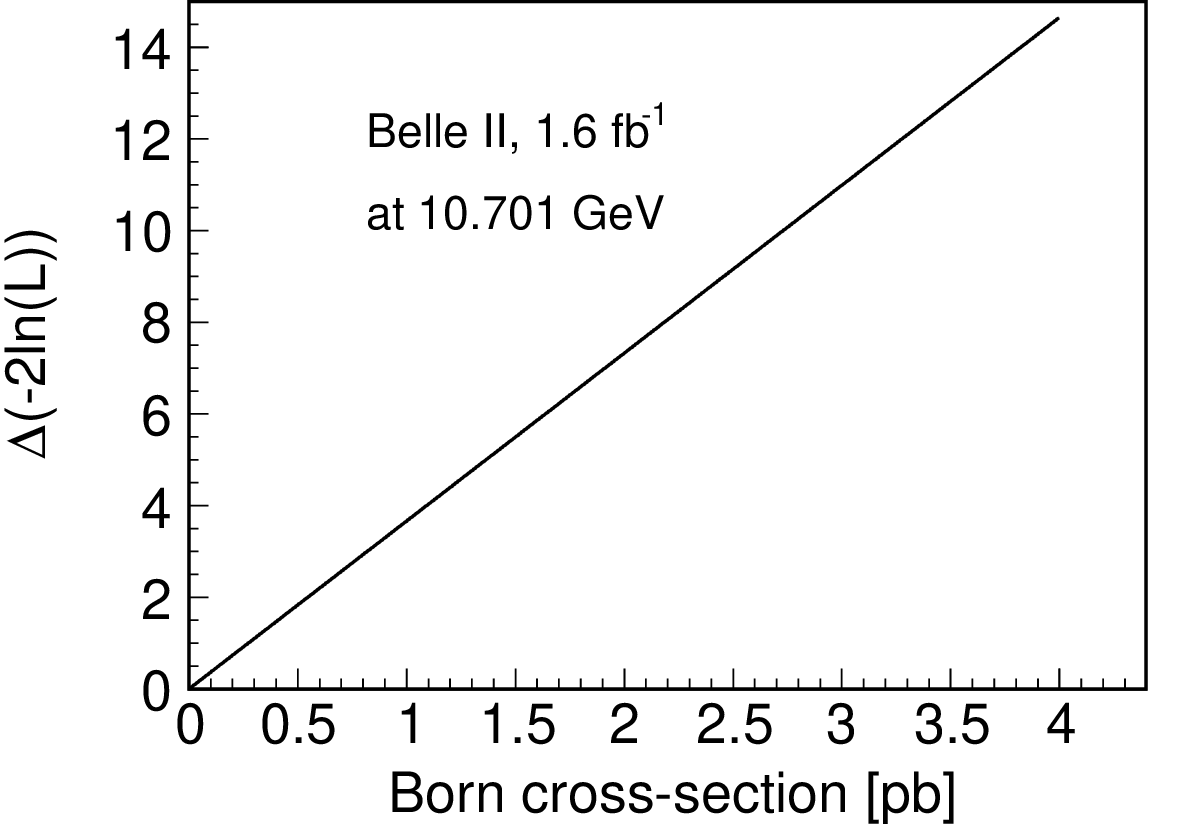}
    \includegraphics[width=0.24\textwidth]{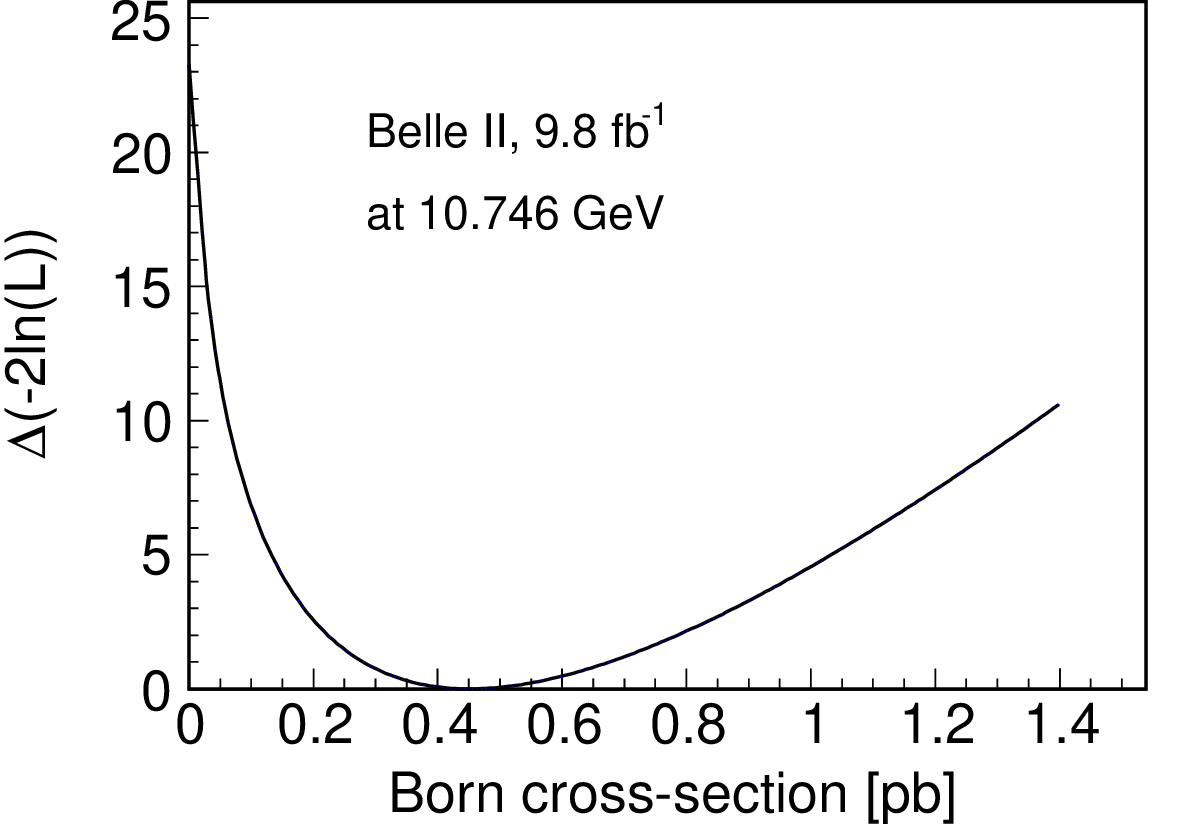}
    \includegraphics[width=0.24\textwidth]{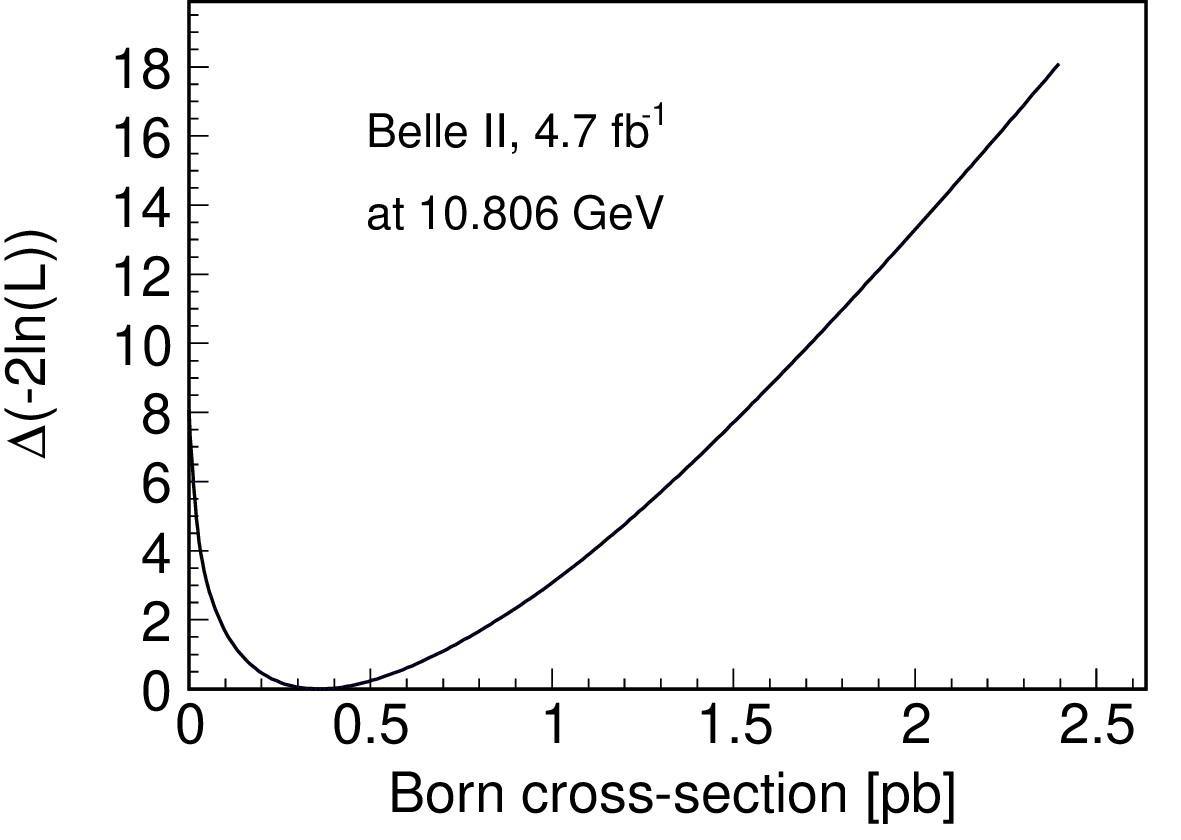}
\\
\end{center}
\caption{$\Delta(-2\ln\mathcal{L})$ as a function of the Born
cross-section in the fit to the $e^+e^-\to\eta\Upsilon(2S)$ signal yield.}\label{fig:fitTo_etaY1S}
\end{figure*} 

\newpage



%

\section{Invariant mass distributions of $\pi^+\pi^-\pi^0$ in $e^+e^-\to\eta\Upsilon(1S)$}

\begin{figure*}[h]
\begin{center}
    \includegraphics[width=0.24\textwidth]{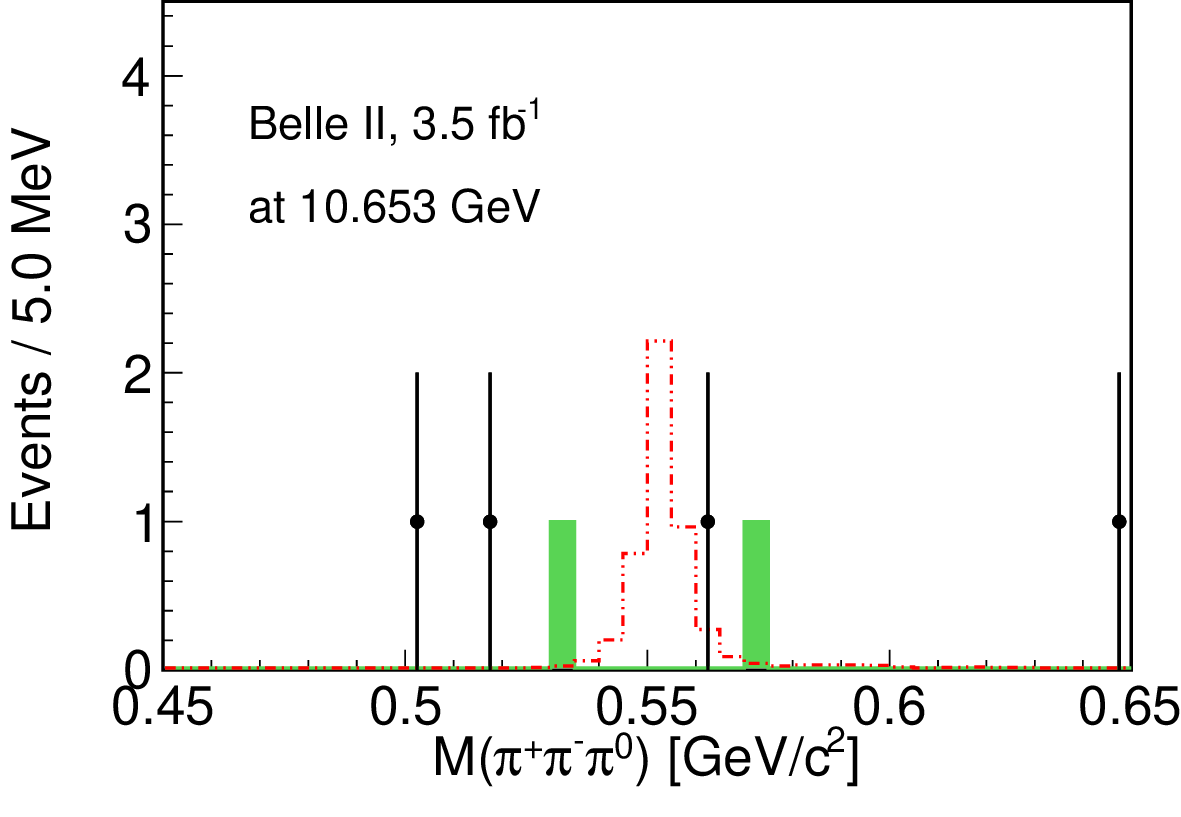}
    \includegraphics[width=0.24\textwidth]{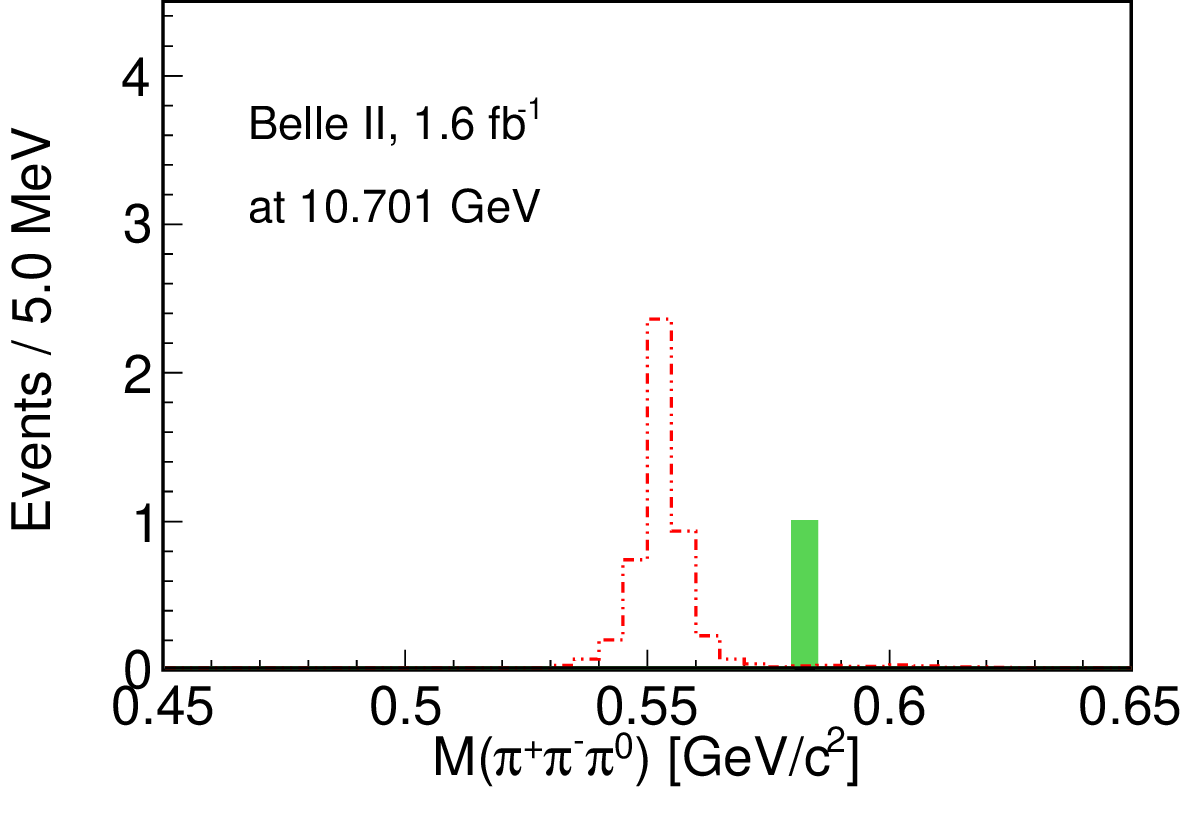}
    \includegraphics[width=0.24\textwidth]{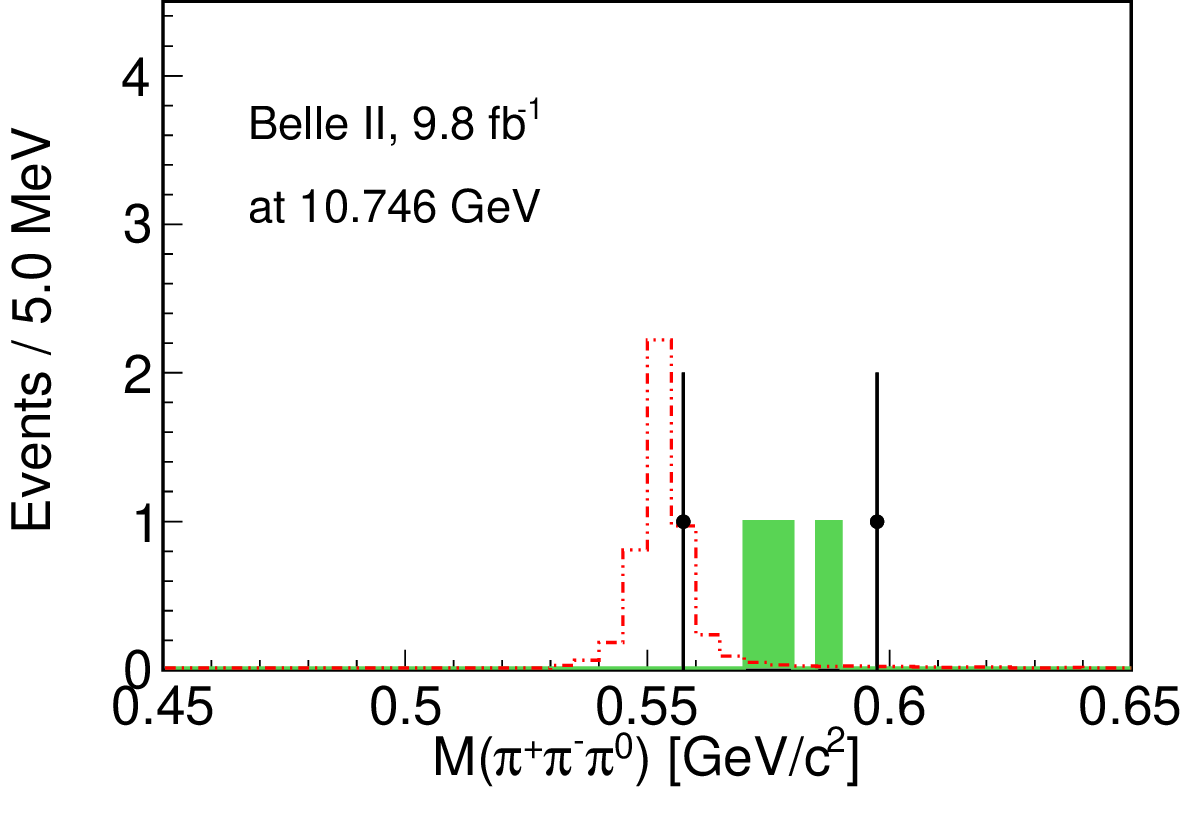}
    \includegraphics[width=0.24\textwidth]{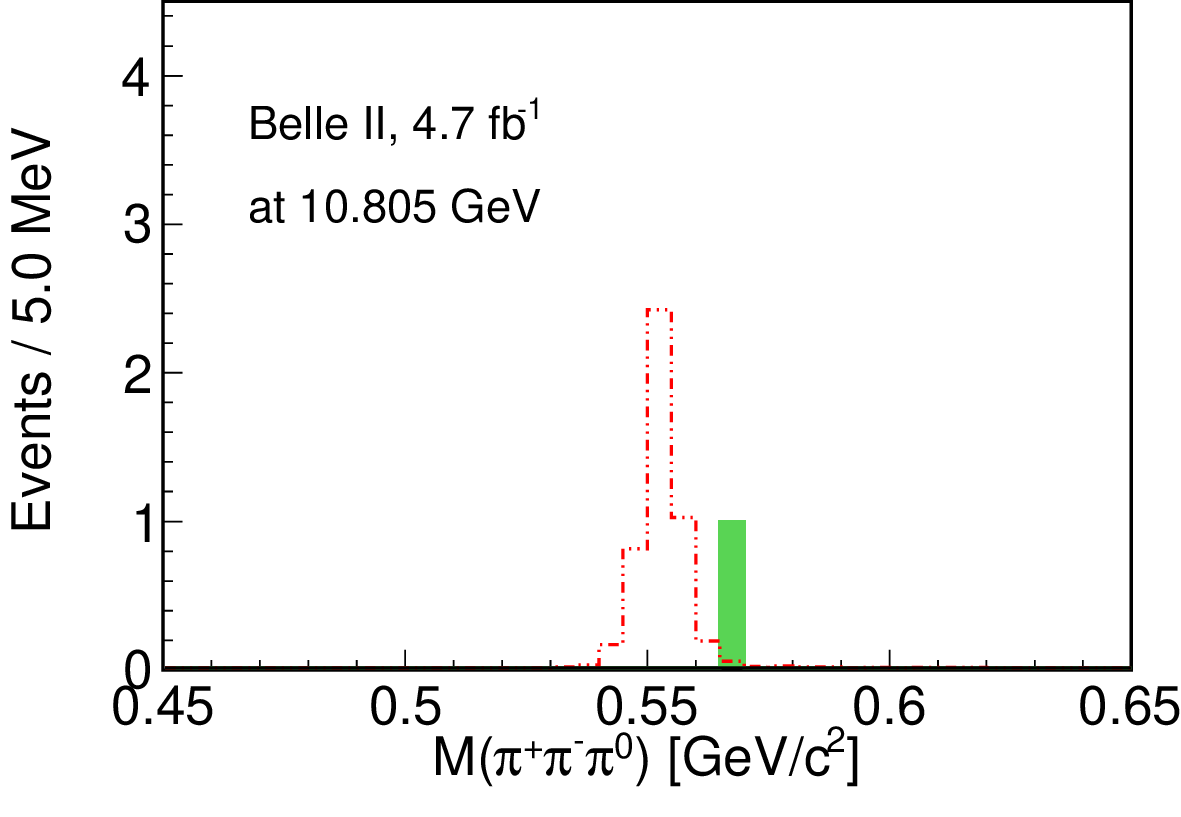}
\\
\end{center}
\caption{Invariant mass distributions of $\pi^+\pi^-\pi^0$ in the $e^+e^-\to\eta\Upsilon(1S)$ channel. Dots with error bars are from data in the $\Upsilon(1S)$ signal region, green shaded histograms are from events in $\Upsilon(1S)$ sideband region, and the red histograms are from the signal MC simulation (with arbitrary normalization).}\label{fig:fitTo_etaY1S}
\end{figure*} 

\section{Invariant mass distributions of $M(\pi^+\pi^-\chi_{bJ})$ in $e^+e^-\to\gamma X_b$ }

\begin{figure*}[ht]
\begin{center}
    \includegraphics[width=0.24\textwidth]{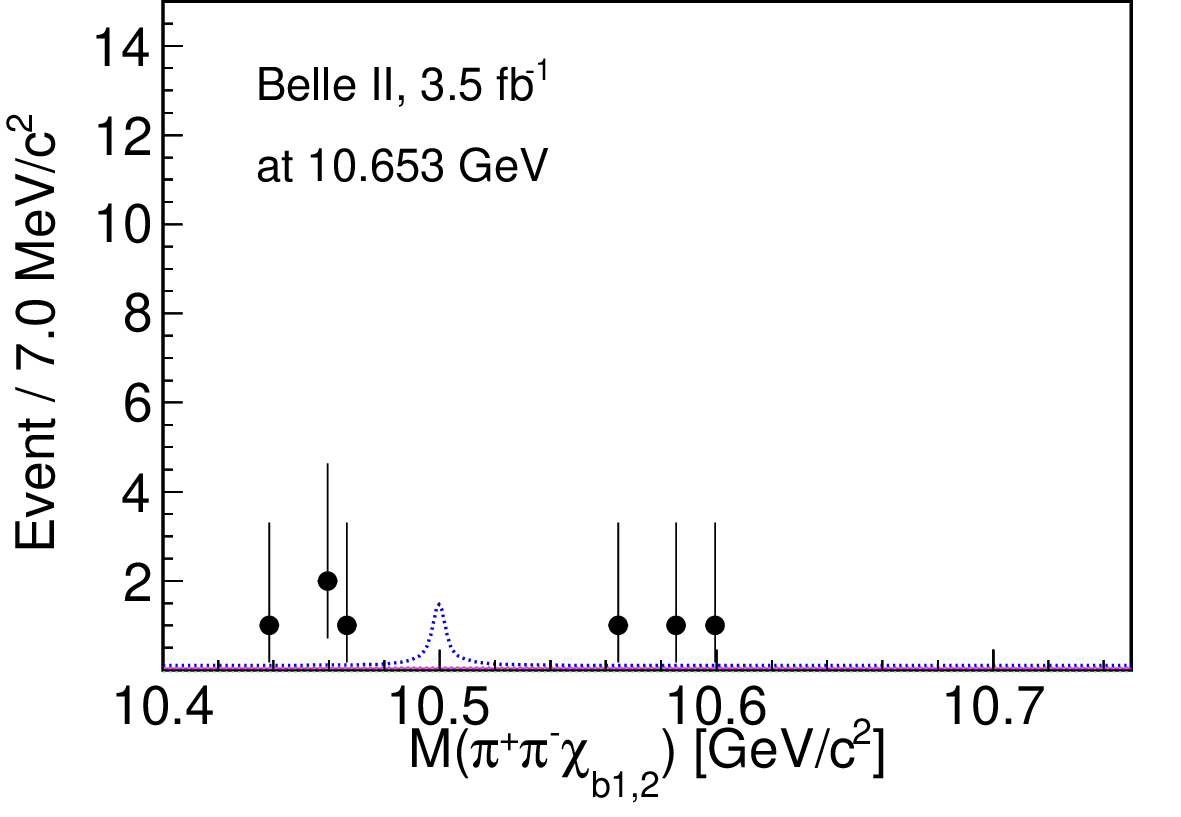}
    \includegraphics[width=0.24\textwidth]{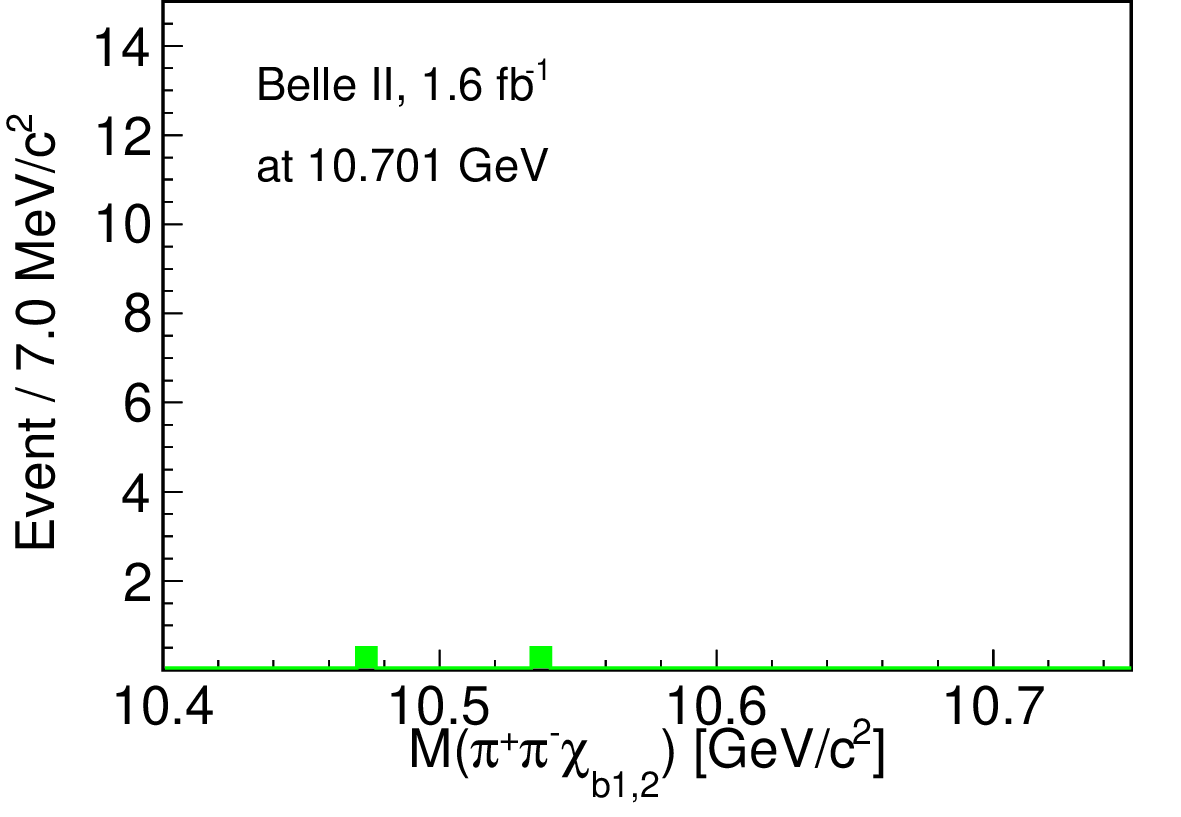}
    \includegraphics[width=0.24\textwidth]{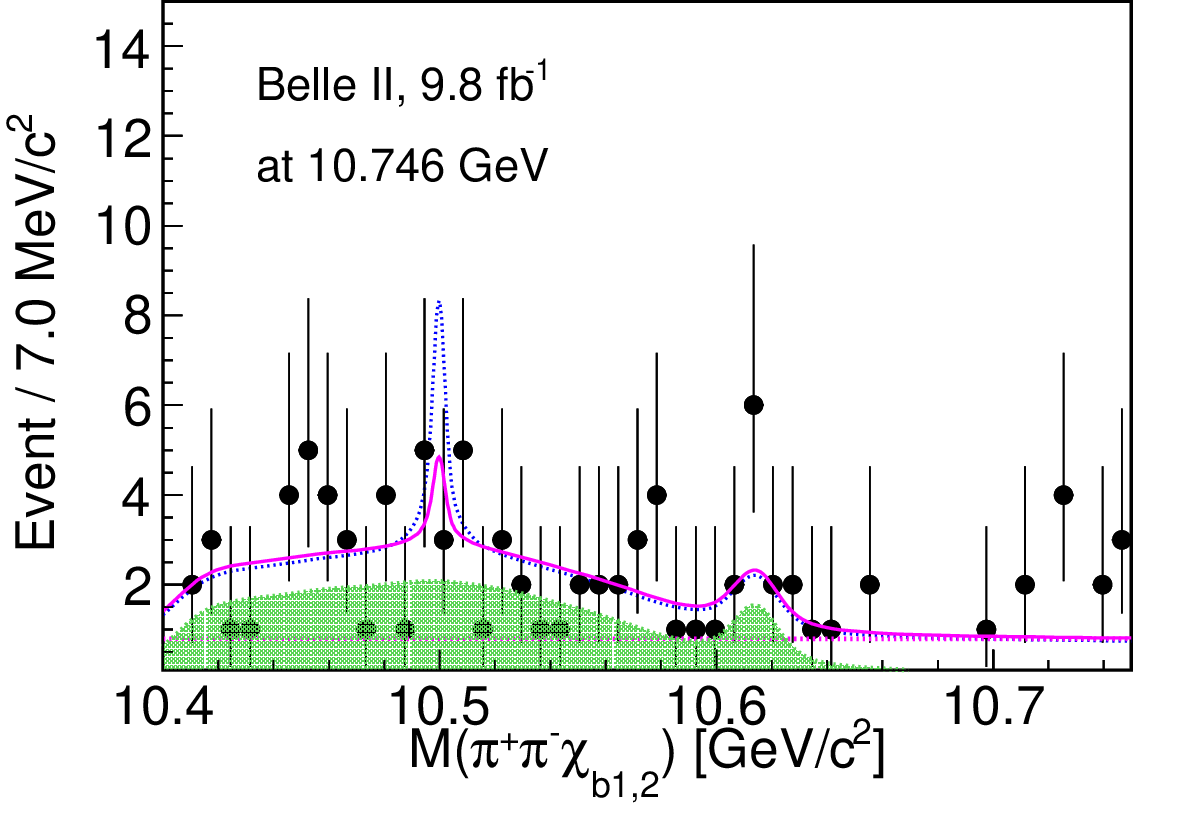}
    \includegraphics[width=0.24\textwidth]{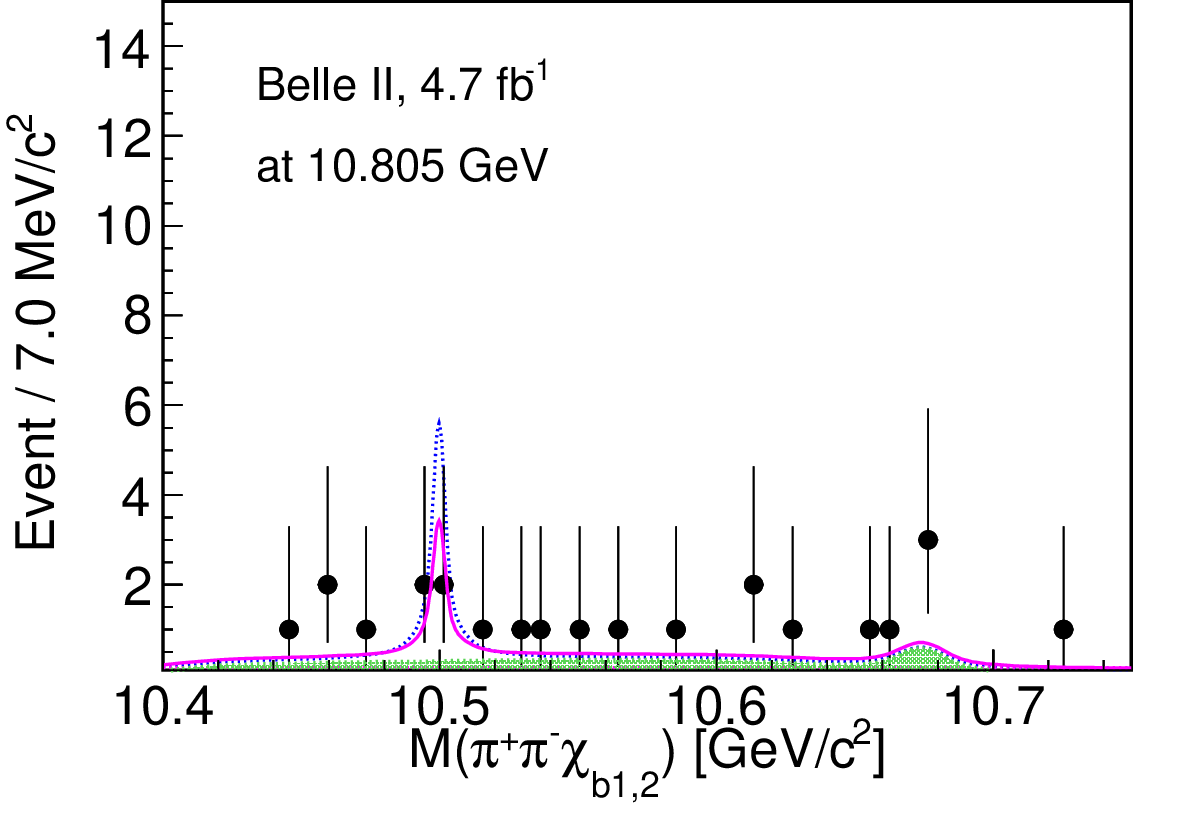}

\end{center}
\caption{Invariant mass distributions of $M(\pi^+\pi^-\chi_{bJ})$ in the $e^+e^-\to\gamma X_b$ channel superimposed with the fit results. Dots with error bars are from data, while the solid line presents the nominal fit result, and the shaded histograms represent the background from $\omega\chi_{bJ}$ and $\pi^+\pi^-\Upsilon(2S)[\gamma\chi_{bJ}]$. Blue dashed lines represent the fit results with the contribution of $X_b$ set to the upper limits at 90\% credibility level .}\label{fig:fitTo_xgamma}
\end{figure*}  

\newpage
\section{Summary of systematic uncertainty}

\begin{table*}[h]
\small
  \begin{center}
\begin{tabular}{@{\extracolsep{\fill}}l c c c c c c c c c c }
      \hline Mode & $\mathcal{B}$ & $\mathcal{L}$ & Tracking & Photon($\pi^0$) & Generator & Trigger &  \,\, Fit \,\,  & \,\,ISR\,\,& \,\,$\mathrm{Model}$\,\, &\,\,{Sum}\,\,  \\
        \hline \multicolumn{2}{l}{$10.653$ GeV} & & & & \\
        \hline
        $\eta \Upsilon(2S)$ & 8.9 & 0.7  & 2.6  &  3.7 &  1.4 & 1.0 &     1.2 & 31.5 &  $<0.1$  & 33.1 \\
        $\eta \Upsilon(1S)$ & 1.9 & 0.7  & 1.0  &  4.0 &  1.4 & 1.0 &   - &  2.0 & $<0.1$  &  5.3  \\

        $\gamma X_b$       & 5.9 & 0.7  & 1.7 &   3.8 &  1.4 & 1.0 &       - & 4.0 & 8.2   & 11.1\\
        \hline \multicolumn{2}{l}{$10.701$ GeV} & & &  &\\
        \hline
        $\eta \Upsilon(2S)$ & 8.9 & 0.7  & 2.4 &  3.7 & 1.4 & 1.0 &      - & - &$<0.1$    &  10.4 \\
        $\eta \Upsilon(1S)$ & 1.9 & 0.7  & 1.0 &  4.1 & 1.4 & 1.0 &    - & - & $<0.1$   & 4.5 \\

        $\gamma X_b$       & 5.9 & 0.7  & 1.7 &  3.8 & 1.4 & 1.0 &       - & - &9.1   &  11.8\\
        \hline
        \multicolumn{2}{l}{$10.746$ GeV} & & & & \\
        \hline
        $\eta \Upsilon(2S)$ & 8.9 & 0.7  & 2.2 & 3.7 & 1.4 & 1.0 &       1.2  &  8.9 &$<0.1$   &  13.5\\
        $\eta \Upsilon(1S)$ & 1.9 & 0.7  & 1.0 & 4.1 & 1.4 & 1.0 &      -  &  2.3 &$<0.1$  & 5.5 \\

        $\gamma X_b$       & 5.9 & 0.7  & 1.7 & 3.8 & 1.4 & 1.0 &         -  &  1.7 &12.2 &  14.4 \\
        \hline
        \multicolumn{2}{l}{$10.805$ GeV} & & &  &\\
        \hline
        $\eta \Upsilon(2S)$ & 8.9 & 0.7  & 2.0 & 3.7 & 1.4 & 1.0 &     6.0 & 8.3 &   $<0.1$    &14.3 \\
        $\eta \Upsilon(1S)$ & 1.9 & 0.7  & 1.0 & 4.1 & 1.4 & 1.0 &     - & 2.3 &$<0.1$  &  5.5  \\

        $\gamma X_b$       & 5.9 & 0.7  & 1.7 & 3.7 & 1.4 & 1.0 &      - & 1.3 & 9.7   &  12.3 \\
      \hline
\end{tabular}
\end{center}
\caption{Summary of the systematic uncertainties in percent for the cross section measurement of $e^+e^-\to\pi^+\pi^-\Upsilon(nS)$ and $e^+e^-\to\gamma X_{b}$. The symbol ``-" denotes an uncertainty value which is not applicable. }
\label{tab:born_sys}
\end{table*}

\end{document}